\documentclass[twocolumn]{aastex631}

\usepackage{amsmath} 

\usepackage{hyperref}
\usepackage{verbatim}

\newcommand\omd{\omega_{\rm dyn}}
\newcommand\omu{\Omega_{\rm U}}
\newcommand\req{R_{\rm eq}}
\newcommand\ru{R_{\rm U}}
\newcommand\pu{P_{\rm U}}
\newcommand\omp{\Omega_{\rm p}}

\newcommand\omplmn{\Omega_{{\rm p},\ell mn}}
\newcommand\degd{{\rm deg\ d}^{-1}}
\newcommand\ajump{\alpha_{\rm jump}}
\newcommand\agrad{\alpha_{\rm grad}}
\newcommand\rhoone{\rho_1}
\newcommand\rcore{r_{\rm core}}
\newcommand\tcore{t_{\rm core}}
\newcommand\taucore{\tau_{\rm core}}
\newcommand\rbreak{r_{\rm break}}

\begin{document}


\shorttitle{Uranian Seismology}
\shortauthors{Mankovich et al.}
\title{Setting the Stage for Uranian Seismology from Rings and Radial Velocities}

\author{Christopher R. Mankovich}\email{mankovich@jpl.nasa.gov}
\author{A. James Friedson}
\author{Marzia Parisi}
\affiliation{Jet Propulsion Laboratory, California Institute of Technology, 4800 Oak Grove Dr., Pasadena, CA 91109, USA}
\author{Stephen Markham}
\affiliation{New Mexico State University Department of Astronomy, Las Cruces, NM 88003, USA}
\author{Janosz W. Dewberry}
\affiliation{Canadian Institute for Theoretical Astrophysics, 60 St George St., Toronto, ON M5S 3H8, Canada}
\author{James Fuller}
\affiliation{TAPIR, Mailcode 350-17, California Institute of Technology, Pasadena, CA 91125, USA}
\author{Matthew M. Hedman}
\affiliation{Department of Physics, University of Idaho, Moscow, ID 83843, USA}
\author{Alex Akins}
\author{Mark D. Hofstadter}
\affiliation{Jet Propulsion Laboratory, California Institute of Technology, 4800 Oak Grove Dr., Pasadena, CA 91109, USA}

\begin{abstract}
    { 
        A Uranus orbiter would be well positioned to detect the planet's free oscillation modes, whose frequencies can resolve questions about Uranus's weakly constrained interior. 
        We calculate the spectra that may manifest in resonances with ring orbits or in Doppler imaging of Uranus's visible surface, using a wide range of interior models that satisfy the present constraints.
        Recent work has shown that Uranus's fundamental (f) and internal gravity (g) modes have appropriate frequencies to resonate with Uranus's narrow rings.
        We show that even a single $\ell=2$ f or g mode detected in ring imaging or occultations can constrain Uranus's core extent and density.
        Fully fluid models typically have $\ell=2-7$ f mode frequencies slightly too high to resonate among the narrow rings.
        If Uranus has a solid core that f modes cannot penetrate, their frequencies are reduced, rendering them more likely to be observed.
        A single $\ell\gtrsim7$ f mode detection would constrain Uranus's unknown rotation period.
        Meanwhile, the different technique of Doppler imaging seismology requires specialized instrumentation but could deliver many detections, with best sensitivity to acoustic (p) modes at mHz frequencies.
        Deviations from uniform frequency spacing can be used to locate density interfaces in Uranus's interior, such as a sharp core boundary.
        Shallower nonadiabaticity and condensation layers complicate this approach, but higher-order frequency differences can be analyzed to disentangle deep and near-surface effects.
        The detection of normal modes by a Uranus orbiter would help to discern among the degenerate solutions permitted by conventional measurements of the planet's static gravity field.
    }
\end{abstract}

\section{Introduction}\label{sec.intro}
Normal mode seismology is a promising means of probing the interiors of giant planets. To date the most direct seismic constraints on giant planet structure have come from ring seismology at Saturn (Kronoseismology; see \citealt{2013AJ....146...12H,2014MNRAS.444.1369H,2016Icar..279...62F,2019Icar..319..599F,2019AJ....157...18H,2021Icar..37014660F,2022PSJ.....3...61H,
1993Icar..106..508M,2014Icar..242..283F,2019ApJ...871....1M,2021NatAs...5.1103M,2021PSJ.....2..198D,2023PSJ.....4...59M}){. I}ndirect evidence of {normal mode oscillations} looms in the accelerations experienced by the Cassini \citep{2020PSJ.....1...27M} and Juno \citep{2022NatCo..13.4632D} spacecraft during their {repeated} close passages of Saturn and Jupiter respectively. 

What of the more distant planets in our solar system? Planets with masses and densities similar to Uranus and Neptune are common among transiting exoplanets (e.g., \citealt{2017AJ....154..108J}), but our most detailed information about {their} prototypes comes from a single Voyager 2 flyby of each planet in the 1980s.
With {the Uranus Orbiter and Probe (UOP)} identified as {the} priority for NASA {Flagship} exploration in the most recent planetary science decadal survey (Origins, Worlds, and Life (OWL); \citealt{NAP26522}), consideration needs to be given to how an orbiter might detect Uranian normal modes to unlock new constraints on the elusive structure of what have conventionally been called the ``ice giants.'' Here we explore the power of two different methodologies to make inferences on Uranus's interior structure based on measurements of normal mode frequencies enabled by an orbiter: ring seismology and Doppler imaging seismology.
{We aim to show how these two independent techniques could be practically applied to open questions about Uranus's interior. Seismology may be the best window into Uranus's hugely uncertain {distribution of chemical elements}, which in turn is intimately tied to fluid stability, thermal and chemical transport processes, rotation state, magnetic field generation, tidal response, and Uranus's formation history. Even putting aside the implications for Uranus's interior, ring seismology may be an essential step toward understanding the dynamical processes sculpting Uranus's unique ring system {(\citealt{1979ApJ...233..857G,1981Natur.292..703C}; see also \citealt{1984prin.conf..513S} for a review of linear spiral wave theory)}. Readying these techniques is therefore critical to answering the key science questions defined as part of the OWL Decadal Strategy.}

In tandem with groundbreaking work on Saturn ring seismology (\citealt{1991Icar...94..420M,1993Icar..106..508M}), \cite{Marley1988UranusAbstract} proposed that normal modes could be responsible for {the radial confinement of} the narrow Uranian rings revealed by Voyager. For all their rich structure, {most regions of} Saturn's main rings {are} roughly uniform in density, and hence are conducive to observation of a spectacular array of spiral waves (see \citealt{2018Icar..312..157T}) induced by resonances with orbiting satellites (e.g., \citealt{1981Natur.292..703C}) or Saturn normal modes (\citealt{2013AJ....146...12H} and aforementioned references). Uranus's rings on the other hand are predominantly narrow and dense {\citep{2018prs..book...93N}}, defying the tendency for collisional scattering to diffuse sharp edges and suggesting the influence of some resonant forcing acting as a confinement mechanism {\citep{1987AJ.....93..724P}}. Confirmed resonances between rings and known moons are few \citep{1987AJ.....93..724P,1991uran.book..327F,2017AJ....154..153C,2024Icar..41115957F}, leaving the mechanism for sustaining $\sim10$ narrow rings a mystery. {Going} beyond the list of named rings, high-phase imaging reveals numerous additional features whose confinement mechanism is also unknown \citep{2021PSJ.....2..107H}. 

Pursuing these ideas, \cite{2022PSJ.....3..194A} showed that fundamental (f) modes or low-order internal gravity (g) modes {of} Uranus are in the appropriate frequency range to resonate among these rings, and a measurement {of} the forcing frequency at one or more of these resonances could be used to distinguish between interior models that otherwise satisfy all {available} constraints (e.g., \citealt{2022PSJ.....3...88M,2023AJ....165...27S,2024A&A...684A.191N,2024A&A...690A.105M,2024arXiv241206010L}). Here we seek to quantify the constraining power of this method by modeling Uranus's mode spectrum for statistical samples of interior models. We consider scenarios in which nonaxisymmetric ring modes are detected, and their driving frequencies and azimuthal wavenumbers $m$ measured, in high-resolution imaging or stellar occultations by an orbiting spacecraft. {We show that a single detection of a low-wavenumber Uranus mode resonating in the rings could eliminate a large fraction of models. We {point to} possible sources of confusion and show the value of obtaining two or more independent constraints in the rings.}

{{O}scillation modes also disturb} the planet{'s visible cloud layers}, yielding radial velocities that {may} be {measurable} by a spatially resolving Doppler imager. A time series of these Doppler images can be used to extract frequencies of normal modes. Existing implementations on the ground make use of interferometry \citep{2024PSJ.....5..100S} or magneto-optical filter designs \citep{2022FrASS...968452S}. 
{If all modes are presumed to have the same energy, t}he radial velocity signal is expected to be dominated by acoustic overtone (p) modes.
{This follows in part from the p modes'} higher frequencies, and hence larger velocity perturbations, than the f, g, interface, or inertial modes {that} reside at lower frequency. 
P modes are trapped sound waves, and hence have special value for discriminating interior structure by virtue of probing the sound speed in the interior. In particular, p modes of the same {spherical harmonic} degree $\ell$ and consecutive radial order $n$ are approximately equally spaced in frequency, but real spectra can contain deviations from this equal spacing that are signatures of jumps or kinks in the adiabatic sound speed profile (e.g., \citealt{2001MNRAS.322...85R}). These modes can hence be powerful diagnostics of composition or phase interfaces in the interior. We take low angular degree ($\ell=1$) p modes as an example to show how a sequence of frequency measurements could be used to locate a density interface in Uranus's interior
{, and how combinations of frequencies of different $\ell$ can be used to separate core and near-surface effects.}

Section~\ref{sec.methods} {describes our} planetary interior modeling, {with} Section~\ref{sec.methods.modes} {giving} a brief primer on the normal modes we discuss. Section~\ref{sec.rings} details the possible observable signatures of Uranian seismicity in the rings and {presents retrievals demonstrating} the constraining power of detecting one or more resonances. Section~\ref{sec.doppler} describes {the unique benefits of} an observed p-mode spectrum that might be accessible from Doppler imaging observations, with particular regard to the location of {composition} interfaces {or phase boundaries} in Uranus's interior. Section~\ref{sec.discussion} discusses our findings and outlook, and we summarize in Section~\ref{sec.conclusion}.

\section{Methods}\label{sec.methods}

\subsection{Uranus interior models}\label{sec.methods.models}

Much recent work has been devoted to exploring the space of Uranus interior structures compatible with the gravity field constraints, both those available now and the improved constraints anticipated from radio tracking of a Uranus orbiter. These modeling efforts can be broadly separated into two categories, those built around physical equations of state (EOSs) on one hand, and `empirical' models informed weakly if at all by an EOS on the other. The latter category \citep{2022PSJ.....3...88M,2022MNRAS.512.3124N,2023AJ....165...27S,2024A&A...684A.191N} prioritizes minimal prior information and a maximally inclusive family of acceptable interior profiles. The former category (e.g., \citealt{2013P&SS...77..143N}) emphasizes compatibility with experimental data and physical interpretability, including direct inferences about composition and temperature structure. EOS-based thermal evolution models (e.g., \citealt{2011ApJ...729...32F}) attempt to reconcile Uranus's surprisingly weak intrinsic flux with the age of the solar system, which appears to require at least one superadiabatic boundary layer in the interior
\citep{2016Icar..275..107N,2019A&A...632A..70S,2021A&A...650A.200S,2020A&A...633A..50V,2021PSJ.....2..222S}.

Unfortunately, the existing body of Uranus models is not readily amenable to detailed seismic analysis. The calculation of adiabatic oscillation modes requires knowledge of the adiabatic sound speed $c_s=(\Gamma_1P/\rho)^{1/2}$ and Brunt-V\"ais\"al\"a (buoyancy) frequency
\begin{equation}
    \label{eq.brunt}
    N^2=\frac{g^2\rho}{P}\left(
        \frac{d\ln\rho}{d\ln P}-\frac{1}{\Gamma_1}
    \right),
\end{equation}
both of which depend on the first adiabatic index $\Gamma_1=\left(\frac{\partial\ln P}{\partial\ln\rho}\right)_s$. (Here $s$ denotes specific entropy and $g=Gm/r^2$.) In principle this is known for EOS-based models, but in practice, EOS sources are adapted and blended and obtaining a reliable $\Gamma_1$ can be a challenge. For empirical models $\Gamma_1$ represents essentially a second unknown profile that is not directly constrained by gravity moments, although density solutions combined with EOSs applied post hoc can give some guidance (e.g., \citealt{2024A&A...684A.191N,2024A&A...690A.105M}). 
{For the sake of readiness for seismology applications, we encourage future giant planet modeling efforts to include $\Gamma_1$ or $N^2$ among their outputs whenever possible.}

Faced with these difficulties, we opt to create a new set of Uranus interior models generally built around polytropic pressure-density relations $P=K\rho^{1+1/n}$. Some reasonable assumptions allow us to obtain well-defined and physically realistic buoyancy and sound speed profiles, while retaining enough flexibility to fit Uranus's zonal gravity harmonics and model the essential features of composition interfaces or gradients.
{By default our models assume a fully fluid interior from the atmosphere to the planetary center, with the exception of the ``rigid core'' variation to be described below.}

We consider two {types of} model for the interior structure.
Before we discuss their differences, a common feature of the two is a break in polytropic index positioned at a radius $r=\rbreak$, motivated by the {inability of any deep interior polytrope to accurately model the lower density, highly compressible outer layers of the planet (see \citealt{2020ApJ...891..109M} and \citealt{2024A&A...690A.105M})}.
For example, a \textit{homogeneous} polytrope satisfying Uranus's mass, radius, spin, and $J_2$ requires $n\approx1.3$ and implies a density at $P=1~{\rm bar}$ of $\rhoone=8\times10^{-4}\ {\rm g\ cm}^{-3}$, overestimating the density implied by Voyager radio occultation data by a factor of approximately 2 \citep{1987JGR....9214987L}.
We hence introduce a break to recover realistic near-surface densities and to avoid {biasing the} gravity moments {or} mode frequencies. We find that 1-bar densities compatible with the Voyager data are achieved for atmosphere polytropic indices $n_{\rm atm}$ in the range $1.5$--$3.5$ {and envelope indices $n_{\rm env}$ typically $\lesssim1$}. 
Density is continuous across the break.
{I}n general the best barotrope to tie to interior models is complicated by the uncertain structure of the CH$_4$ condensation layer and abyssal abundance \citep{2011Icar..215..292S,2021PSJ.....2..146M}, as well as latitudinal and temporal variations in the temperature near 1 bar \citep{2020AJ....159...45R}; these will feed into the error on the 1-bar density used to constrain the models. 

The uncertain structure of the deeper H$_2$O condensation zone $(P\gtrsim100$ bar) is another topic of major interest \citep{2017Icar..297..160F,2017A&A...598A..98L,2021PSJ.....2..146M}. Simplified models like the ones we use here could be generalized to account for water condensation and an associated radiative zone, and seismology may prove to be a useful probe of these phenomena.
{In this limit the vertical scale of the water cloud is expected to be small (\citealt{2021PSJ.....2..146M} estimate 10-100 m). Here the nearly discontinuous change in composition and temperature would induce p mode frequency shifts functionally similar to those induced by the artificial envelope/atmosphere break introduced above, the results of which are discussed in detail in Section~\ref{sec.doppler}. Typical f modes and the deep-seated g or interface modes discussed in Section~\ref{sec.rings} would be less sensitive to such a sharp, shallow layer. A stably stratified water cloud may also host its own internal gravity waves, but the negligible amount of mass involved makes their observation unlikely. Hence, our conclusions are not altered by our choice not to model the H$_2$O cloud region explicitly.
}

For the deeper interior structure, we consider two possibilities:

\noindent\textbf{Interface model}: a composite polytrope defined by outer and inner polytropic indices $n_{\rm env}$ and $n_{\rm core}$, joined at a double mesh point at $r=\rcore$ at which the density increases by a factor of $\ajump>1$. 

We choose to set the adiabatic index $\Gamma_1$ equal to the polytropic derivative $\gamma=\frac{d\ln P}{d\ln\rho}=1+1/n$ in each of the two layers, yielding an adiabatic stratification $N^2=0$ within those layers. The intervening discontinuities in $\rho$ and $c_s^2$ modify the spectrum of f and p modes and also introduce interface modes{, which stem from the gravity-wave response to small displacements of an interface between two fluid layers of differing densities} \citep{1976Icar...27..109V}. {These are distinct from g modes, which are internal gravity modes that require a continuous\footnote{Or pseudo-continuous; see \citealt{2015MNRAS.452.2700B}.} stably stratified medium with finite radial extent.}

\noindent\textbf{Gradient model}: similarly to~\cite{2014Icar..242..283F}, we start with {a simple} reference polytrope {of index $n_{\rm env}$} that satisfies Uranus's mass and radius, designate the region $0<r<r_i$ as the core region, and enhance the density there by a constant factor $\agrad>1$. The envelope density outside $r=r_o$ is unchanged{, and $n_{\rm env}$ hence corresponds to the final envelope polytropic index}. Then $\rho(r)$ in the intervening region $r_i<r<r_o$ is set by a linear function 
\begin{equation}
    \rho(r)=\rho_o+\left(\rho_i-\rho_o\right)\left(\frac{r-r_o}{r_i-r_o}\right),\quad r_i<r<r_o,
\end{equation}
where $\rho_o=\rho_{n_{\rm env}}(r_o)$ and $\rho_i=\agrad\,\rho_{n_{\rm env}}(r_i)$ are the densities at the outside and inside of the gradient region. 

Similar to the interface model, we {set $\Gamma_1=\left(\frac{\partial\ln P}{\partial\ln\rho}\right)_s$} to $\frac{d\ln P}{d\ln\rho}$ to guarantee an adiabatic stratification $N^2=0$ outside the gradient region. Inside the gradient region, we choose 
\begin{equation}
    c_s^2=\Gamma_1P/\rho=c_{s,o}^2+(c_{s,i}^2-c_{s,o}^2)\left(\frac{r-r_o}{r_i-r_o}\right),\quad r_i<r<r_o.
\end{equation}
This choice yields positive values of $N^2$ in the gradient region for the models considered here, introducing {a spectrum of} g modes. 

\begin{figure}
    \begin{center}
        \includegraphics[width=\columnwidth]{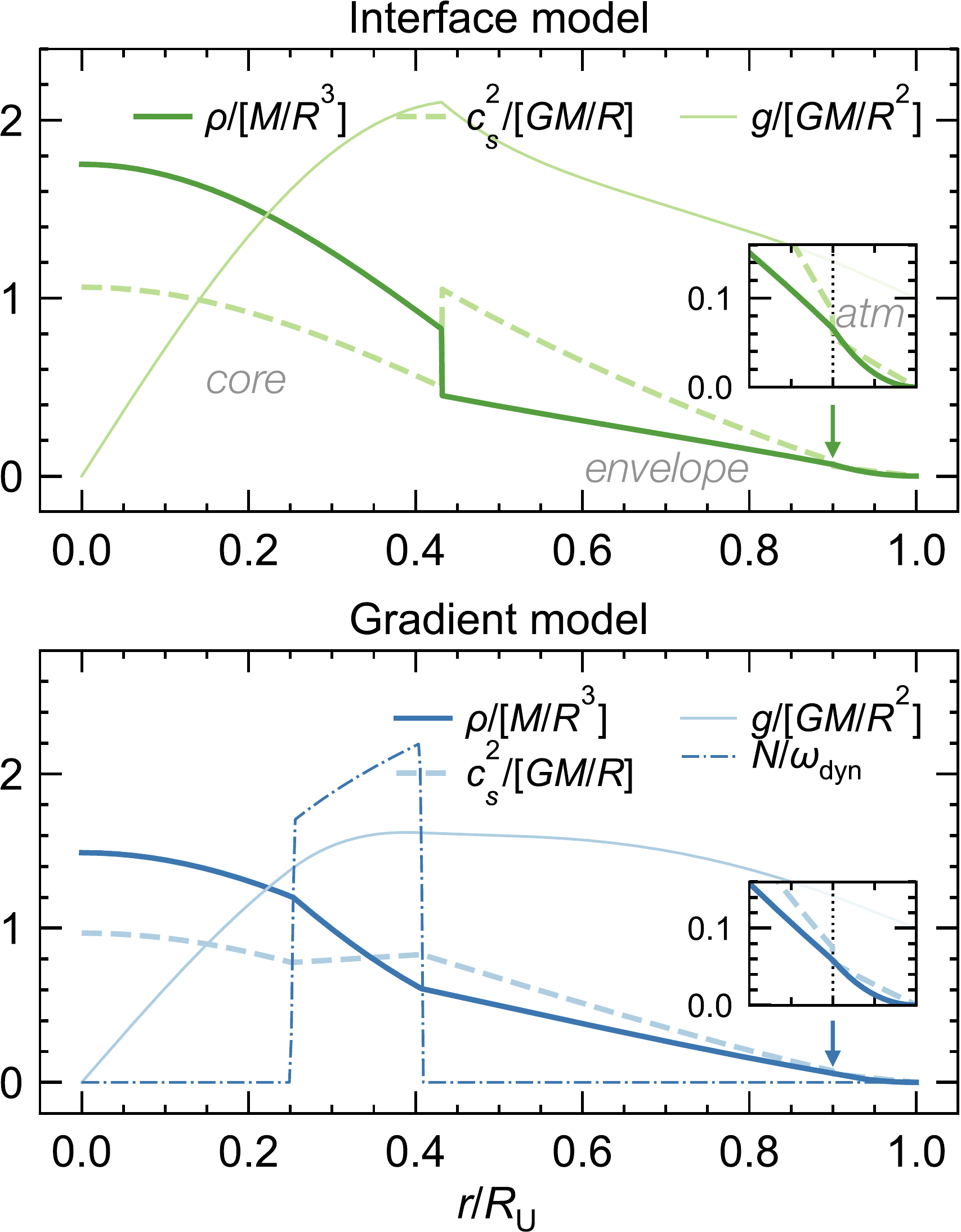}
        \caption{
            Examples of the two types of interior structures considered in this study, one (top) fully adiabatic with a discontinuous core density enhancement and the other (bottom) with a gradual core density enhancement and superadiabaticity in the gradient region.
            Both models have moderate central density (9.{3} and 7.9 g cm$^{-3}$ respectively) among their parent samples.
            The {insets} in each panel {emphasize the density and sound speed structure near} the break between envelope and atmosphere polytropes.
            \label{fig.model_examples}
        }
    \end{center}
\end{figure}
        
In the gradient model, the polytropic constant 
\begin{equation}\label{eq.k_env}
    K_{\rm env}=P/\rho^{1+1/n_{\rm env}}
\end{equation} 
of the underlying reference polytrope is adjusted {during structure} iterations such that the model converges to the desired total mass $M=M_{\rm U}$ to a specified tolerance{.} A similar procedure is used in the interface model, where pressure continuity at the interface relates the two polytropic constants ($K_{\rm env},K_{\rm core}$) and $K_{\rm core}$ is adjusted to achieve Uranus's total mass. In both model types $n_{\rm env}$ is adjusted to satisfy the target value of $J_2${, which varies from one model to the next for reasons we will describe}. The atmosphere and envelope constants $K_{\rm atm}$ and $K_{\rm env}$ are linked by the condition of density continuity at the break $r=\rbreak$.

{The frequencies of the spheroidal oscillation modes of interest generally scale with Uranus's dynamical frequency
\begin{equation}
    \label{eq.dynamical_frequency}
    \omega_{\rm dyn}=\left(\frac{GM_{\rm U}}{\ru^3}\right)^{1/2}.
\end{equation}
Hence, we choose a total mass tolerance $10^{-5}\ M_{\rm U}$ to yield a fractional frequency error less than 1\%, safely smaller than the error introduced by our approximations concerning rotation (see Section~\ref{sec.methods.modes} below).}

{To give an overview of the models we consider, Figure~\ref{fig.model_examples} plots the profiles of $\rho$, $c_s^2$, $g$, and $N$ for just one example of each model class.}
{$N$ is omitted for the interface model, where by construction $N=0$ everywhere except a single point at the interface. Here the application of jump conditions (Section~\ref{sec.methods.modes} below) obviates the need for explicit calculation of $N$ across the discontinuity.}

As a subset of the interface models, we also address the possibility of an idealized, inelastic, shear-free core by enforcing a zero-displacement boundary condition at the interface radius $r=\rcore$ in the interface models. This allows us to assess the major modifications that a frozen core {(e.g., \citealt{2021PSJ.....2..222S})} makes to the spectrum of oscillations in Uranus's fluid envelope, but neglects the nonzero but poorly constrained bulk modulus and shear modulus of the solid core. In reality pressure and shear modes of a solid core may interact with fluid oscillations in the envelope, providing additional seismic diagnostics (see, e.g., \citealt{2014Icar..231...34F}).

We account for rotation by using the theory of figures {(ToF)} \citep{2017A&A...606A.139N,2021PSJ.....2..241N} {to fourth order} to iteratively calculate a self-consistent planetary shape and potential in a rigidly rotating Uranus.
{Each iteration enforces the prescription for $\rho(r)$, $c_s^2(r)$ described above where $r$ corresponds to the mean radii of the isopotential surfaces. These radii} are uniformly scaled such that the outermost zone's equatorial radius matches $\req=25,559$ km. Models commonly assume a deep spin period $\pu=17.24\ {\rm h}$ {guided} by periods derived from the magnetosphere \citep{1986Natur.322...42D}, but Uranus's rotation rate is highly uncertain (e.g., \citealt{2010Icar..210..446H}) and better information may await shape, gravity, and hopefully normal mode measurements by an orbiter. We treat $\pu$ as a free parameter in the range 16--18 h.

\subsubsection{Constraining the density profile}\label{sec.methods.models.constraints}

After Uranus's mass and equatorial radius, the primary constraint on interior structure is $J_2$, which we fit directly by adjusting one of the parameters during ToF iterations (similar to, e.g., \citealt{2023PSJ.....4...95M}). {In all models t}his is accomplished by adjusting $n_{\rm env}${.}
Since the wind-induced part of $J_2$ is not known a priori, each ToF model is in fact adjusted to match {a nuisance sampled parameter} $J_2^{\rm rigid}$, and the total $J_n=J_n^{\rm rigid}+J_n^{\rm winds}$ are compared to data.

Uranus's (and Neptune's) rapid cloud-level winds are thought to be relatively shallow, extending at most several percent into the planet \citep{2013Natur.497..344K}, but are still an important contribution to even the low-order zonal harmonics $J_2$ and $J_4$. We calculate the effect that these jet streams have on the zonal gravity field by solving a thermo-gravitational wind {equation (TGWE) model} as in \cite{2023PSJ.....4...59M}. 
We treat Uranus's winds assuming the 1-bar level has wind speeds consistent with Voyager 2 and Hubble data (per \citealt{2023AJ....165...27S}'s fit; their Equation 16) and apply a simple exponential decay as a function of radial distance from the 1-bar surface.
This {decay function} introduces the $e$-folding depth $d$.
{To spare the computational cost of TGWE calculations for hundreds of thousands of models, we instead precompute such models for a few thousand randomly sampled rigidly rotating models fit to Uranus's $J_2$ and 1-bar density. Polynomial fits for $J_{2n}^{\rm winds}$ as a function of $d$ and $\pu$ enable an efficient comparison to data during the sampling process. Further details are given in Appendix~\ref{app.tgwe}.}

Until a {spacecraft} orbits Uranus, the best constraints on the planet's gravity field come from the precession rates of the rings, as measured from a combination of Voyager 2 and Earth-based occultations \citep{1988Icar...73..349F,2014AJ....148...76J}. More recently \cite{2024Icar..41115957F} made new {measurements} of $J_2$ and $J_4$, accounting for several sources of systematic uncertainty that had not yet been considered. Nonetheless, their adopted solution is statistically compatible with the larger $J_2-J_4$ error ellipse of Jacobson's adopted solution. This work compares models to the centroid of Jacobson's estimates of the coefficients, considering a range of uncertainties. It is instructive to frame the difference between Jacobson and French et al.'s $J_2$ centroids in terms of properties of the interior model: in the interface model from Figure~\ref{fig.model_examples} (top) the $1.39$ ppm difference between Jacobson's and French et al.'s $J_2$ value can be compensated by adjusting the envelope polytropic index $n_{\rm env}$ from $0.5435$ to $0.5437$, changing the model's central density by $\lesssim0.03\%$. The resulting differences in mode frequencies are insubstantial for our purposes.

A Uranus orbiter could improve our knowledge of the zonal gravity moments by orders of magnitude. \cite{2024PSJ.....5..116P} estimate the precision on {the even and odd harmonics from }$J_2$ {to} $J_{10}$ that may be attainable for practical spacecraft orbits. 
Hence, we {also perform retrievals} in which we replace baseline \cite{2014AJ....148...76J} uncertainties with those predicted by \cite{2024PSJ.....5..116P} {in} their NO OCC SAT-LIKE Trajectory 1 case. This relatively conservative case includes 8 pericenters outside the rings; similar precision would be reached by an orbit with pericenter passages inside the rings but limited to 4 to 5 gravity orbits.

Returning to the constraint on density near the surface, recall that both model types feature a `softer' polytrope to describe the atmosphere, allowing the models to attain realistic 1-bar densities at the expense of introducing two parameters $\rbreak$ and $n_{\rm atm}$. To implement the 1-bar density constraint, we ignore the details of 
{shallow atmosphere condensation} 
and consider an ideal gas at the temperature and number density at 1 bar from the nominal atmosphere model of \cite{1987JGR....9214987L}, where an ideal gas with a CH$_4$ mixing ratio between $0$ and $0.04$ has a density $\rhoone=3.647$ to $4.516\times10^{-4}$ g cm$^{-3}$, a range of $24\%$. Reducing the helium to hydrogen mixing ratio from 15/85 to 11/85 (e.g., \citealt{2011Icar..215..292S}) or increasing the temperature by 5 K modulates the density by $\lesssim4\%$, an {e}ffect overwhelmed by the uncertain abyssal CH$_4$ abundance. Erring on the side of permissiveness, we assign models a Gaussian likelihood in $\rhoone$ centered on $3.647\times10^{-4}\ {\rm g\ cm}^{-3}$ with standard deviation $10^{-4}\ {\rm g\ cm}^{-3}$.

\begin{table*}[]
    \begin{center}
    \refstepcounter{table}
    \textbf{Table \thetable} \\
    Interior model parameters \\
\begin{tabular}{llll}
Interface model                     &                                                                       &                                               &                                                   \\ \hline
\multicolumn{1}{l}{Parameter}       & \multicolumn{1}{l}{Meaning}                                           & \multicolumn{1}{l}{Permitted range}           & \multicolumn{1}{l}{Comment}                       \\ \hline
\multicolumn{1}{l}{$n_{\rm atm}$}   & \multicolumn{1}{l}{Atmosphere polytropic index}                       & \multicolumn{1}{l}{$(0.5, 3.5)$}              & \multicolumn{1}{l}{MCMC}                          \\ 
\multicolumn{1}{l}{$n_{\rm env}$}   & \multicolumn{1}{l}{Envelope polytropic index}                         & \multicolumn{1}{l}{$(0.5, 3.5)$}              & \multicolumn{1}{l}{Adjusted to fit $J_2$}         \\ 
\multicolumn{1}{l}{$n_{\rm core}$}  & \multicolumn{1}{l}{Core polytropic index}                             & \multicolumn{1}{l}{-}                         & \multicolumn{1}{l}{Fixed at 1}                    \\ 
\multicolumn{1}{l}{$\rbreak$}       & \multicolumn{1}{l}{Atmosphere/envelope break radius$^{\rm a}$}                  & \multicolumn{1}{l}{-}                         & \multicolumn{1}{l}{Fixed at 0.9}                  \\ 
\multicolumn{1}{l}{$\rcore$}        & \multicolumn{1}{l}{Envelope/core jump radius$^{\rm a}$}                         & \multicolumn{1}{l}{$(0.1, 0.8)$}              & \multicolumn{1}{l}{MCMC}                          \\ 
\multicolumn{1}{l}{$\ajump$}        & \multicolumn{1}{l}{Core density enhancement factor}                   & \multicolumn{1}{l}{$(1, 3)$}                  & \multicolumn{1}{l}{MCMC}                          \\ 
\multicolumn{1}{l}{$K_{\rm core}$}  & \multicolumn{1}{l}{Core polytropic proportionality constant}          & \multicolumn{1}{l}{Any}                       & \multicolumn{1}{l}{Adjusted to fit total mass}    \\
\multicolumn{1}{l}{$d$}             & \multicolumn{1}{l}{Decay depth of jet streams, units $\req$}          & \multicolumn{1}{l}{$(0.01, 0.10)$}            & \multicolumn{1}{l}{MCMC}                          \\
\multicolumn{1}{l}{$\pu$}           & \multicolumn{1}{l}{Bulk rotation period, h}                           & \multicolumn{1}{l}{$(16, 18)$}                & \multicolumn{1}{l}{MCMC}                          \\  
                                    &                                                                       &                                               &                                                   \\
Gradient model                  &                                                                           &                                &                                                                  \\ \hline
\multicolumn{1}{l}{Parameter}       & \multicolumn{1}{l}{Meaning}                                           & \multicolumn{1}{l}{Permitted range}           & \multicolumn{1}{l}{Comment}                       \\ \hline
\multicolumn{1}{l}{$n_{\rm atm}$}   & \multicolumn{1}{l}{Atmosphere polytropic index}                       & \multicolumn{1}{l}{$(0.5, 3.5)$}              & \multicolumn{1}{l}{MCMC}                          \\ 
\multicolumn{1}{l}{$n_{\rm env}$}   & \multicolumn{1}{l}{Reference (and final envelope) polytropic index}   & \multicolumn{1}{l}{$(0.5, 3.5)$}              & \multicolumn{1}{l}{Adjusted to fit target $J_2$}  \\ 
\multicolumn{1}{l}{$\rbreak$}       & \multicolumn{1}{l}{Atmosphere/envelope break radius$^{\rm a}$}                  & \multicolumn{1}{l}{-}                         & \multicolumn{1}{l}{Fixed at 0.9}                  \\ 
\multicolumn{1}{l}{$r_o$}           & \multicolumn{1}{l}{Gradient outer boundary radius$^{\rm a}$}                    & \multicolumn{1}{l}{$(r_i, r_{\rm break})$}    & \multicolumn{1}{l}{MCMC}                          \\ 
\multicolumn{1}{l}{$r_i$}           & \multicolumn{1}{l}{Gradient inner boundary radius$^{\rm a}$}                    & \multicolumn{1}{l}{$(0.1, 0.8)$}              & \multicolumn{1}{l}{MCMC}                          \\ 
\multicolumn{1}{l}{$\agrad$}        & \multicolumn{1}{l}{Core density enhancement factor}                   & \multicolumn{1}{l}{$(1, 3)$}                  & \multicolumn{1}{l}{MCMC}                          \\ 
\multicolumn{1}{l}{$K_{\rm env}$}   & \multicolumn{1}{l}{Reference polytrope proportionality constant}      & \multicolumn{1}{l}{Any}                       & \multicolumn{1}{l}{Adjusted to fit total mass}    \\
\multicolumn{1}{l}{$d$}             & \multicolumn{1}{l}{Decay depth of jet streams, units $\req$}          & \multicolumn{1}{l}{$(0.01, 0.10)$}            & \multicolumn{1}{l}{MCMC}                          \\
\multicolumn{1}{l}{$\pu$}           & \multicolumn{1}{l}{Bulk rotation period, h}                           & \multicolumn{1}{l}{$(16, 18)$}                & \multicolumn{1}{l}{MCMC}                          \\  
\end{tabular}
\caption{
    \label{tab.parameters}
    \footnotesize{
$^{\rm a}$ These quantities are specified as a fraction of the volumetric mean radius $R$ of the $P=1$ bar surface, which varies along with the oblateness from one model to the next. With $\req=25,559$ km a constant, $R$ is hence controlled mostly by the assumed spin period $\pu$.
}
}
\end{center}
\end{table*}

Table~\ref{tab.parameters} summarizes the free parameters in each model class, states their role as an adjusted or sampled parameter, and gives the uniform prior volume allowed for each. {The sampling process, described in more detail below, is very similar to that used by \cite{2023PSJ.....4...59M},} but fit {here} to a mere three data points $J_2$, $J_4$, and $\rhoone$.

\begin{figure}
    \begin{center}
        \includegraphics[width=\columnwidth]{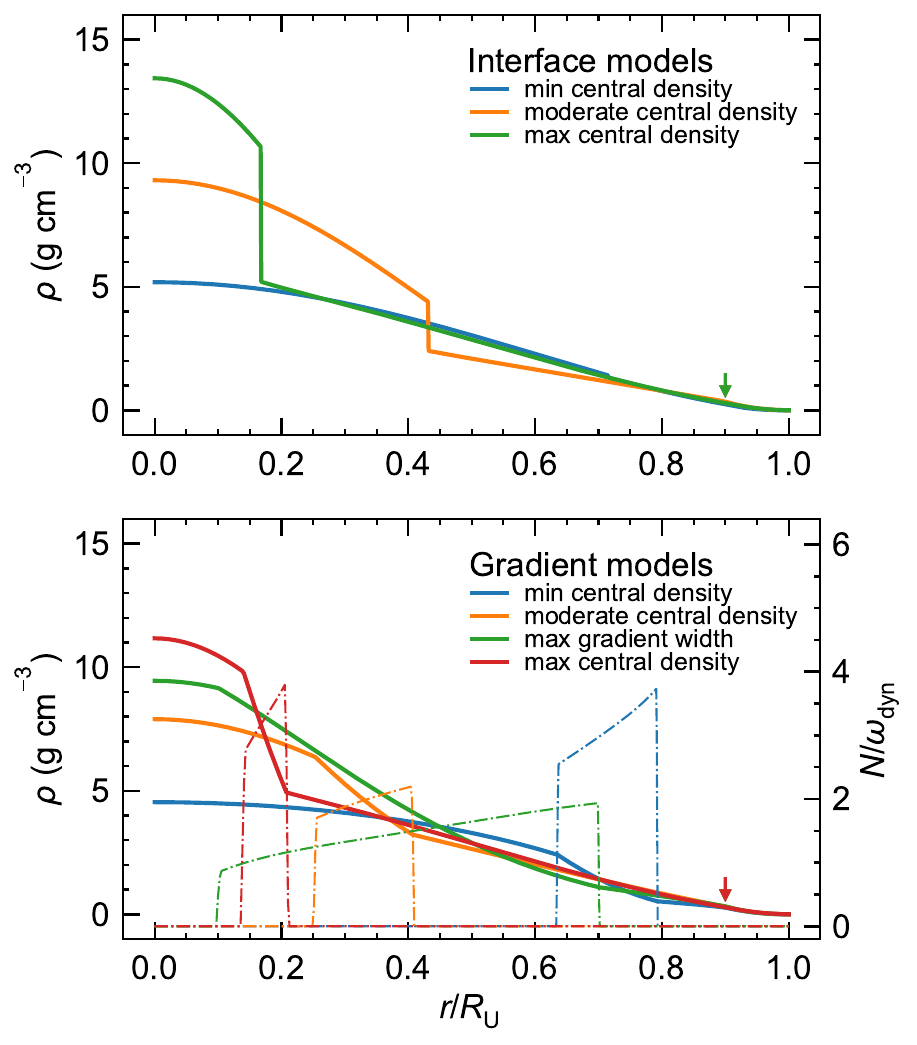}
        \caption{
            \label{fig.models_of_interest}
            As in Figure~\ref{fig.model_examples}, but {emphasizing the range of models permitted by $J_2$, $J_4$, and $\rhoone$. These} representative models {are} used for detailed mode calculations. 
            \textit{Top:} Interface models.
            \textit{Bottom:} Gradient models. Brunt-V\"ais\"al\"a frequency (dot-dashed, read on right axis) is plotted as a {fraction} of Uranus's dynamical frequency (Equation~\ref{eq.dynamical_frequency}).
            {Arrows mark the transition to the atmospheric polytrope.}
        }
    \end{center}
\end{figure}

For each family of models, we select representative well-fitting end-member models for detailed seismic modeling{. T}heir $\rho$ and $N$ profiles are shown in Figure~\ref{fig.models_of_interest}. 
For interface models these are the models with minimum, moderate, and maximum central density with respect to their parent distribution. Here ``moderate'' refers to the midpoint of a quantity's minimum and maximum. Among the gradient models we select minimum/moderate/maximum central density and maximum gradient width $r_o-r_i$.
Given the small number of data being fit for our gravity-only samples, we naturally recover a highly degenerate posterior probability distribution. We aim to show how one or more seismic measurements can aid in clearing up these degeneracies.

\subsection{Oscillation modes}\label{sec.methods.modes}
We solve the fourth-order system of ordinary differential equations describing the adiabatic oscillation modes using GYRE\footnote{\url{https://gyre.readthedocs.io/en/v7.1}} \citep{2013MNRAS.435.3406T}. Modes are obtained in the absence of rotation in an equivalent spherical model defined on the isopotential mean level surfaces of the oblate ToF model. 
In this limit of slow rotation, the angular structure of each normal mode is simply given by a single spherical harmonic $Y_\ell^m$, where the integers $\ell=0,\ldots,\infty$ and $m=-\ell,\ldots,\ell$ are called the angular degree and azimuthal order. However, many normal modes can exist with the same $\ell$ and $m$, and uniquely specifying a mode requires a third integer $n$ that counts the number of nodes as a function of radius in the mode's eigenfunction. This radial order differentiates between the broad categories of g, f, and p modes that we will describe shortly.
For brevity we follow \citep{2023PSJ.....4...59M} in referring to individual modes using the notation $_\ell^m\cdot_n$ so that, e.g., the $\ell=7$, $m=5$, $n=2$ p mode would be labeled as $_7^5$p$_2$
. We label the interface modes as $_\ell^m$i, not to be confused with inertial modes {(see \citealt{DewberryLai2022,2023Icar..40515711F})}, which {are limited to lower frequencies $\lesssim2\omu$ and generally yield weaker gravity and radial velocity perturbations per unit mode energy compared to the modes} considered in this paper.
{We adopt the convention that frequency is always positive, taking $m>0$ ($m<0$) to label prograde (retrograde) propagating modes with respect to Uranus's rotation.}

Normal mode amplitudes are unknown in Uranus. 
{A variety of excitation processes are known in stars, including feedback processes assocated with the radiative opacity or fusion rate ($\kappa$ and $\epsilon$ mechanisms), tidal forcing (heartbeat stars), and stochastic forcing by surface convection as in the Sun.}
Our best information about the amplitude spectrum of normal modes in a giant planet comes from Saturn, where the amplitudes of ring waves excited by Saturn can be used to constrain the amplitudes of the perturbing modes \citep{2014MNRAS.444.1369H,afgibo}. The {amplitude} spectrum is complicated and no prevailing theory exists for the excitation and dissipation processes at work in Saturn, much less Jupiter, Uranus, or Neptune{, but deep atmospheric rock storms \citep{2018Icar..306..200M} and giant impacts \citep{2019ApJ...881..142W} are promising candidates for excitation}. {W}e refrain from any detailed modeling of the amplitude spectrum, and work with modes normalized to unity mode inertia \citep{2010aste.book.....A}. In this normalization the energy of the $\ell mn$ mode is simply $\frac12\omega_{0,\ell mn}^2$. {Lacking a theory for mode excitation in Uranus, we can nevertheless estimate the mode amplitude required to generate ring forcing potentials comparable to those already measured from satellite-driven ring resonances \citep{2024Icar..41115957F}. We show in Appendix~\ref{app.ring_mode_energy} that Uranus f modes can generate observable ring signatures if their nondimensional amplitudes are of the same order of magnitude as the Saturn f modes responsible for the waves observed in the C ring.}

The influence of rigid rotation on oscillation frequencies is included using the first-order Coriolis coefficient 
\begin{equation}
    \label{eq.beta}
    \beta_{\ell mn} = 1 - \frac{\int_0^R\left(2\xi_{\rm r}\xi_{\rm h}+\xi_{\rm h}^2\right)\,\rho r^2\,dr}{\int_0^R\left(\xi_{\rm r}^2+\ell[\ell+1]\xi_{\rm h}^2\right)\,\rho r^2\,dr}
\end{equation}
{derived from perturbation theory (\citealt{1989nos..book.....U}; here $\xi_{\rm r}$ and $\xi_{\rm h}$ are the radial and horizontal displacement eigenfunctions.)} 
Working in this approximation, the final inertial frame frequency of mode $\ell mn$ is
\begin{equation}
    \begin{split}
    \label{eq.inertial_frame_frequency}
    \sigma_{\ell mn} &= \omega_{\ell mn}+m\omu \\
    &= \omega_{0,\ell mn}+m\beta_{\ell mn}\omu
    \end{split}
\end{equation}
where $\omega_{\ell mn}$ is the frequency in the frame co-rotating with Uranus and $\omega_{0,\ell mn}$ is the frequency obtained in the absence of rotation.
In contrast to Saturn, where rapid rotation means that $\mathcal O(\Omega^2)$ terms contribute up to $(\Omega/\omd)^2=14\%$ to mode frequencies, at Uranus's more modest rotation the second order terms contribute $\lesssim3\%$. More accurate treatments of the influence of rigid and differential rotation have been established (e.g., \citealt{2006A&A...455..621R,2021PSJ.....2..198D,2022MNRAS.516..358D}) but are not warranted at this exploratory stage. 

{The differential rotation responsible for Uranus's atmospheric jet streams (see \citealt{2005Icar..179..459S}) modifies the mode frequencies. The magnitude of these perturbations is set by the radial extent of the differentially rotating layers, and the spatial structure of the mode in question. Frequency shifts are likely largest for sectoral ($\ell=|m|$) modes due to their equatorially concentrated eigenfunctions, which would tend to coherently sample the retograde flow surrounding Uranus's equator. (See \citealt{2023PSJ.....4...59M} for first-order rotation kernels in Saturn f modes, where sectoral modes predominantly sample that planet's prograde equatorial jet.) Tesseral ($\ell>|m|$) mode eigenfunctions tend to extend to higher latitudes, inviting self-cancellation as they sample both prograde- and retrograde-rotating latitudes. In any case, as noted by \cite{2022PSJ.....3..194A}, the small frequency shifts induced by Uranus's atmospheric flows are overwhelmed by the uncertainties that follow from Uranus's highly uncertain spin period. For present purposes it is therefore appropriate to neglect the influence of differential rotation on the oscillation modes.}

Double mesh points are included to correctly treat the density interface (in models where one exists) and the break in polytropic index, which introduces a {discontinuity} in $\Gamma_1$. At these locations GYRE enforces jump conditions to guarantee continuity of the radial displacement and Lagrangian pressure perturbation.
{In models with rigid (non-oscillatory) cores, zero radial displacement is enforced at core boundary; otherwise, GYRE applies its default inner boundary condition imposing regularity of the eigenfunctions.}
{At the surface of the model, GYRE applies boundary conditions on the Eulerian gravitational potential perturbation and the Lagrangian pressure perturbation that follow from the vanishing of the background density.
Strictly speaking, this condition is violated by our truncation of the atmosphere polytrope at $P=10^{-2}$ bar where $\rho\sim10^{-5}\ {\rm g\ cm}^{-3}$. The near-surface behavior of the eigenfunctions changes somewhat for different truncation pressures, but we find that the influence on f, g, and p mode frequencies is minor for our purposes.}

What follows is a brief introduction to the types of oscillation mode considered here.

\noindent\textbf{F modes.} The fundamental modes manifest the fact that if the planetary surface is perturbed locally, gravity acts as a restoring force and the disturbance propagates away as a wave. F modes are hence surface gravity waves somewhat similar to deep water waves. Their amplitudes decay approximately exponentially below the planetary surface. However, at low $\ell$ this decay scale is comparable to the planetary radius, lending these modes sensitivity to regions close to the planetary center. The f mode frequencies approximately obey \citep{2010aste.book.....A}
\begin{equation}
    \label{eq.f_mode_frequency}
    \omega_f^2=\frac{g_{\rm surf}\sqrt{\ell(\ell+1)}}{R}=\omega_{\rm dyn}^2\sqrt{\ell(\ell+1)}
\end{equation}
with $g_{\rm surf}$ the gravity at the planetary surface.
Compared to other types of modes, the radial phase coherence of f~modes (no nodes as a function of radius; $n=0$) means that they produce intrinsically larger gravity perturbations for a fixed mode energy.

\noindent\textbf{Interface modes.} Similarly, interface modes have maximum amplitude at a density interface and decay approximately exponentially on either side. Their frequencies $\omega_i$ approximately obey \citep{2010aste.book.....A}
\begin{equation}
    \label{eq.interface_mode_frequency}
    \omega_i^2 = \frac{g_i\sqrt{\ell(\ell+1)}}{r_i}\left(\frac{\rho_i-\rho_o}{\rho_i+\rho_o}\right),
\end{equation}
where $g_i$ is the gravity at the interface and $\rho_o$, $\rho_i=\ajump\rho_o$ are the density at either side of the interface as described in Section~\ref{sec.methods.models}. A comparison of~\ref{eq.f_mode_frequency} and \ref{eq.interface_mode_frequency} suggests that f modes can be cast as a special case of interface mode that happens to be bounded on one side by a vacuum. 
{Thanks to the rapid decay of their eigenfunctions with distance from the interface, prospects for observing these modes from outside the planet diminish steeply as the interface is located deeper. However, coupling with modes with larger surface expressions (e.g., the f modes) can make the interface modes easier to observe, as shown in Section~\ref{sec.rings.avoided} below.}

\noindent\textbf{G modes.} These are internal gravity waves restored by buoyancy; their existence hence requires stable stratification $N^2>0$ somewhere in the interior. Formally, g modes are wavelike where $0<\omega_g<N$ and evanescent {elsewhere}. But proximity to f modes, in terms of frequency or of eigenfunction overlap, can yield modes of mixed f and g mode character, as has been argued to take place in Saturn (e.g., \citealt{2014Icar..242..283F}).

\noindent\textbf{P modes.} These are oscillatory in regions where $\omega^2>S_\ell^2$, with the Lamb frequency $S_\ell$ given by
\begin{equation}
    \label{eq.lamb}
    S_\ell^2=\frac{\ell(\ell+1)c_s^2}{r^2}.
\end{equation}
P modes hence propagate in the envelope outside a characteristic turning point {where $\omega^2=S_\ell^2$, at which point refraction from the increasing sound speed causes the wavevector to become purely horizontal and rays turn back toward the planetary surface.}

A characteristic frequency of the p mode spectrum is the ``large frequency separation''
\begin{equation}
    \label{eq.delta_nu}
    \Delta\nu = \left(2\int_0^R c_s^{-1}dr\right)^{-1}
\end{equation}
which gives the constant frequency spacing between p modes of the same degree $\ell$ and consecutive radial order \citep{1980ApJS...43..469T}. This constant spacing applies in the limit of high radial order ($n_p\gg1$), but real spectra exhibit deviations that can be used to probe internal structure (e.g., \citealt{2002RvMP...74.1073C,2013ARA&A..51..353C}) as we will show for Uranus.

\begin{figure}
    \begin{center}
        \includegraphics[width=\columnwidth]{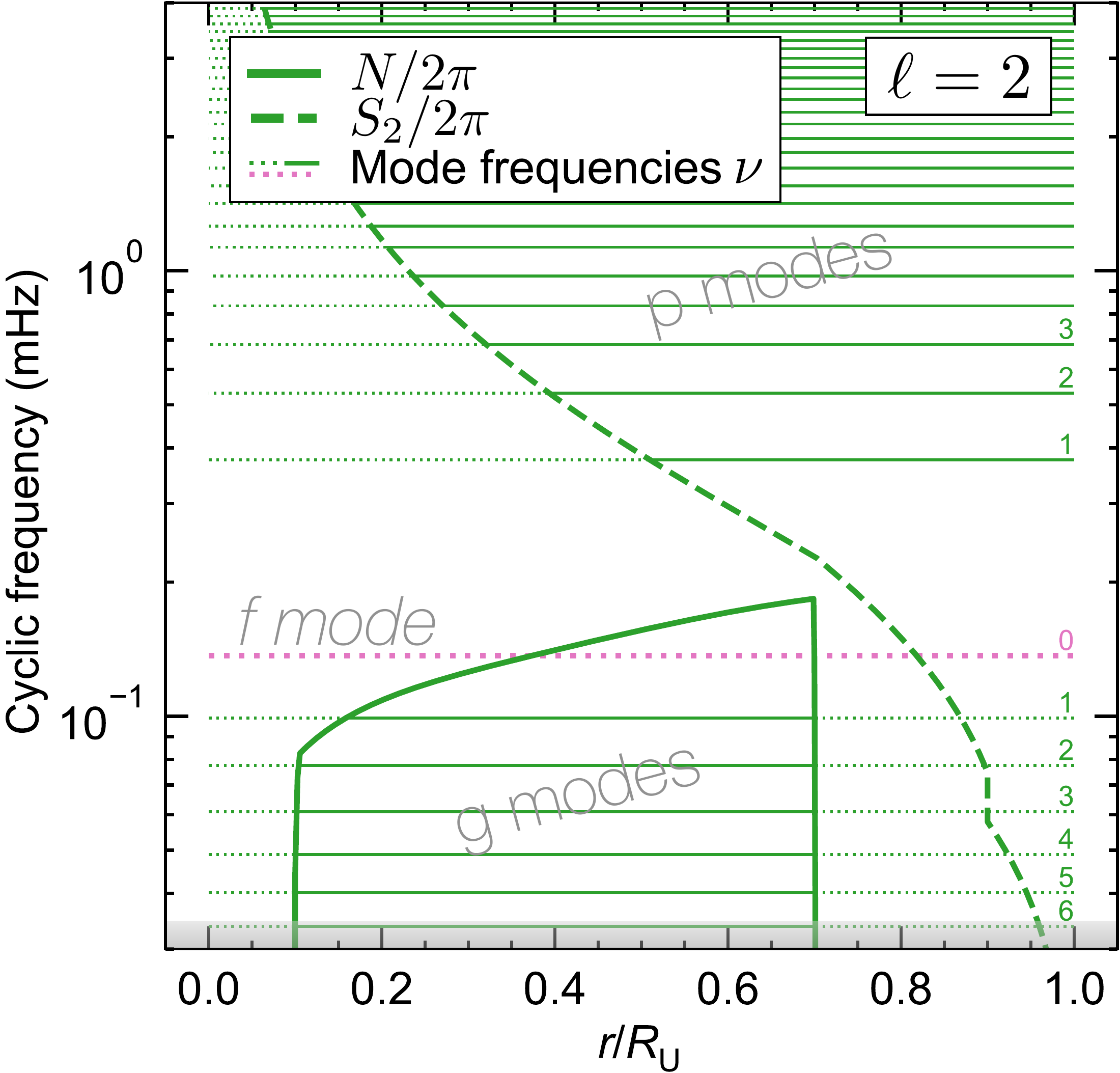}
        \caption{
            \label{fig.propagation_diagram}
            An $\ell=2$ propagation diagram for the gradient model with maximum gradient width.
            P modes are oscillatory where their frequencies are greater than the Lamb frequency (thick dashed curve; see Equation~\ref{eq.lamb}).
            G modes are oscillatory where their frequencies are less than the Brunt-V\"ais\"al\"a frequency (thick solid curve; see Equation~\ref{eq.brunt}).
            Horizontal lines are calculated {$m=0$} mode frequencies for the model.
            {Dotted} regions are evanescent zones where mode amplitudes decay with distance from their wave propagation region but are still generally nonzero.
            {Integer labels give the p or g mode radial order.}
            {The grey shaded region at low frequency indicates the inertial frequency range $\omega<2\Omega_{\rm U}$ in which the Coriolis force becomes an important restoring force and our approximations regarding rotation (Section~\ref{sec.methods.modes}) are no longer valid.}
        }
    \end{center}
\end{figure}

Figure~\ref{fig.propagation_diagram} shows a propagation diagram, a useful atlas of how mode frequency dictates the behavior of the oscillation in the interior. {We plot the $\ell=2$, $m=0$ spectrum for the `max gradient width' model from the gradient sample, for which the defining frequencies $N$ and $S_2$ are shown as a function of radius. The fact that $N<S_2$ throughout the interior enforces a strict frequency hierarchy between g, f, and p modes in this case. Models with larger values of $N$ (e.g., those with more abrupt composition gradients) could lead to g-mode spectra that overlap with f- and even p-mode frequencies. This possibility becomes more remote toward higher $\ell$ where f- and p-mode frequencies increase while the g modes remain confined to $\nu\lesssim N$. A similar diagram for one of our interface models would lack a g mode spectrum but feature a single interface mode. Its frequency may be less than or comparable to the f mode frequency or even low-order p mode frequencies, depending on the properties of the interface (see Equation~\ref{eq.interface_mode_frequency} and Section~\ref{sec.rings.avoided}).}

We ignore all lower frequency modes, particularly inertial regime ($\omega<2\omu$) modes. This category includes Rossby and other inertial modes; these generally have smaller intrinsic gravitational perturbations than f/g/interface modes at fixed mode energy and so may be less amenable to detection in the rings, although they may have been detected in Saturn's C ring (\citealt{2022PSJ.....3...61H}; see \citealt{2023Icar..40515711F}). These modes are of too low frequency and radial velocity amplitude to be tractable in Doppler imaging seismology.

\subsection{Parameter estimation}\label{sec.methods.parameter_estimation}
We sample each model's 6- or 7-dimensional parameter space using a Markov chain Monte Carlo (MCMC) framework similar to the one documented in \cite{2021NatAs...5.1103M} and \cite{2023PSJ.....4...59M}.
At the core of this approach is the sampling tool \texttt{emcee}\footnote{\url{https://emcee.readthedocs.io/en/v3.1.4}} \citep{2013PSP..125..306F}.
{Table~\ref{tab.parameters} summarizes the model parameters.}
Models are evaluated using a multivariate Gaussian likelihood in $J_2$, $J_4$, and $\rhoone$ as described {in Section~\ref{sec.methods.models}}. 
Samples presented in Section~\ref{sec.rings} further consider the addition of pattern speeds of hypothetical ring resonances. 
{For a more detailed description of the sampling process we refer to Appendix~\ref{app.parameter_estimation}.}

\begin{figure*}[h]
    \begin{center}
        \includegraphics[width=0.9\textwidth]{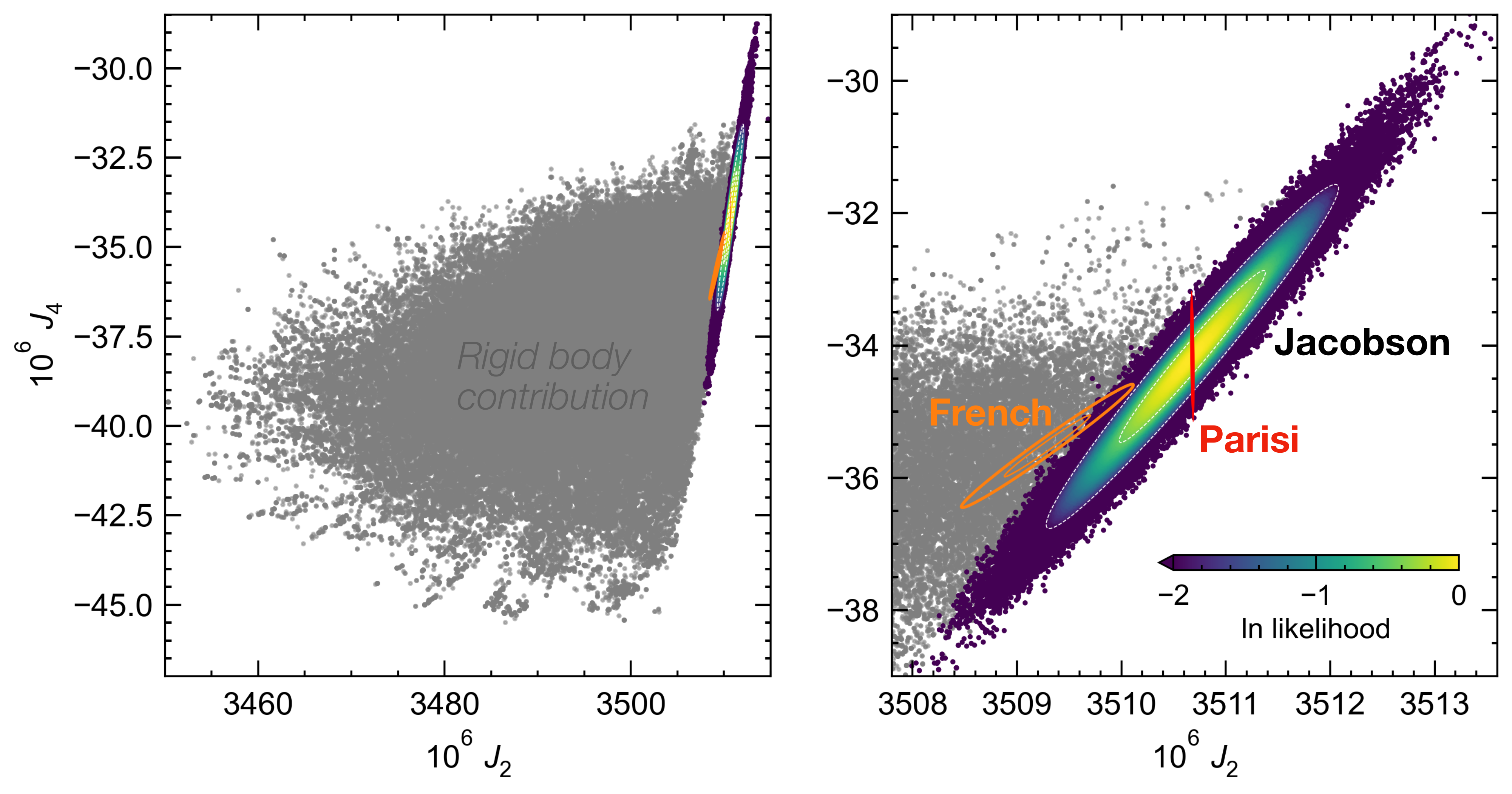}
        \caption{
            \label{fig.j2n_scatter_gradient_break}
            {Low degree z}onal gravity moments for gradient type models of Uranus.
            {(Interface models yield a similar diagram.)}
            The grey points show the part of the $J_{2n}$ arising from the rigidly rotating background structure.
            The colorful points show the total including the addition of $\Delta J_{2n}^{\rm winds}$; these are colored according to their likelihood with low to high likelihood running purple to green to yellow.
            The sample is fit to \cite{2014AJ....148...76J} gravity ($1\sigma$ and $2\sigma$ ellipses in dashed white). 
            Shown for comparison are the more recent \cite{2024Icar..41115957F} measurements ($1\sigma$ and $2\sigma$ ellipses in orange) and improved uncertainties anticipated by \cite{2024PSJ.....5..116P} for radio tracking of a future Uranus orbiter (narrow red $1\sigma$ ellipse; see text).
        }
    \end{center}
\end{figure*}

Figure~\ref{fig.j2n_scatter_gradient_break} shows our baseline sample of gradient models within the context of current and future $J_2$ and $J_4$ measurements. 
Note that a wide distribution of interior structures are compatible with the data. 
{Uranus's interior is poorly constrained,} manifested here as the broad distribution of rigid body $J_{2n}$ (grey points){.}
Figure~\ref{fig.corner_gradient_break} in the appendices shows the highly degenerate space of interior model parameters in detail.

\section{Ring seismology: f, g, and interface modes}\label{sec.rings}

In the rings, we focus on the search for outer Lindblad resonances (OLRs){ with Uranus's prograde oscillation modes. Lindblad resonances are locations} where ring orbits experience periodic forcing at a frequency commensurate with the radial epicyclic frequency {\citep{1979ApJ...233..857G}}.
OLRs are expected for typical prograde planetary modes, for which the forcing potential rotates faster than the ring mean motion. 
In contrast, resonances with satellites usually produce inner Lindblad resonances (ILRs) due to the forcing potential rotating more slowly than the ring mean motion.
{(ILRs with planet modes are also possible but are limited to higher $m$ values; these are revisited below.)}
Each prograde Uranus mode with inertial frequency given by Equation~\ref{eq.inertial_frame_frequency} has a pattern speed $\omplmn=\sigma_{\ell mn}/m$ which, fed into the OLR condition 
(see \citealt{2019ApJ...871....1M} for the form we solve)
yields a unique location in the rings. 
Note that for a given forcing frequency, the resonance location depends on the zonal gravity moments through the epicyclic frequency. {T}he \cite{2014AJ....148...76J} and \cite{2024Icar..41115957F} gravity solutions yield consistent OLR locations to within $\approx 30$ m.

\begin{figure*}[h]
    \begin{center}
        \includegraphics[width=0.7\textwidth]{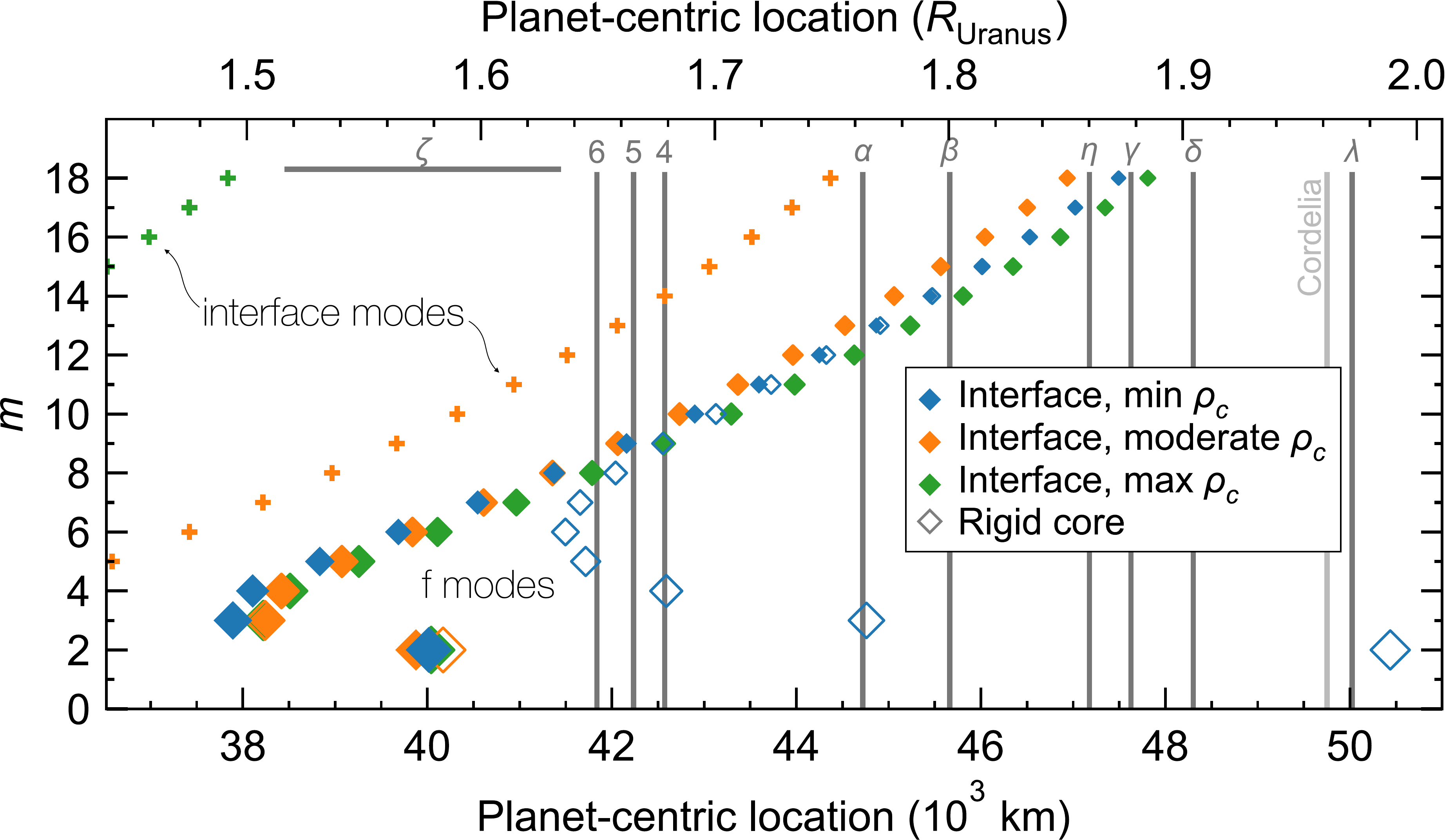}
        \caption{
            \label{fig.ring_olrs_interface}
            Outer Lindblad resonance locations amid Uranus's rings for $\ell=m$ modes in a series of interface-based models compatible with all available data.
            {Colors map to interior model as in Figure~\ref{fig.models_of_interest}; see legend.}
            {F m}odes with larger gravitational potentials at the planetary surface are rendered as larger points.
            Open symbols show f mode OLRs for models with a perfectly rigid core.
            Larger rigid cores push the {low degree} f modes to lower frequencies and hence their OLRs to greater orbital distances.
            {Crosses denote the interface modes, shown with uniform point size for clarity. 
            The interface modes (crosses) in the model with the largest core (min $\rho_c$; blue points) are too low frequency to resonate in the rings.} 
            In the model with the most compact core (max $\rho_c$; green points), the rigid and fluid core cases have indistinguishable spectra.
            {Note that the $_2^2$f mode in the min $\rho_c$ fluid core model (blue) nearly coincides with that in the max $\rho_c$ model (green).}
        }
    \end{center}
\end{figure*}

\begin{figure*}[h]
    \begin{center}
        \includegraphics[width=0.7\textwidth]{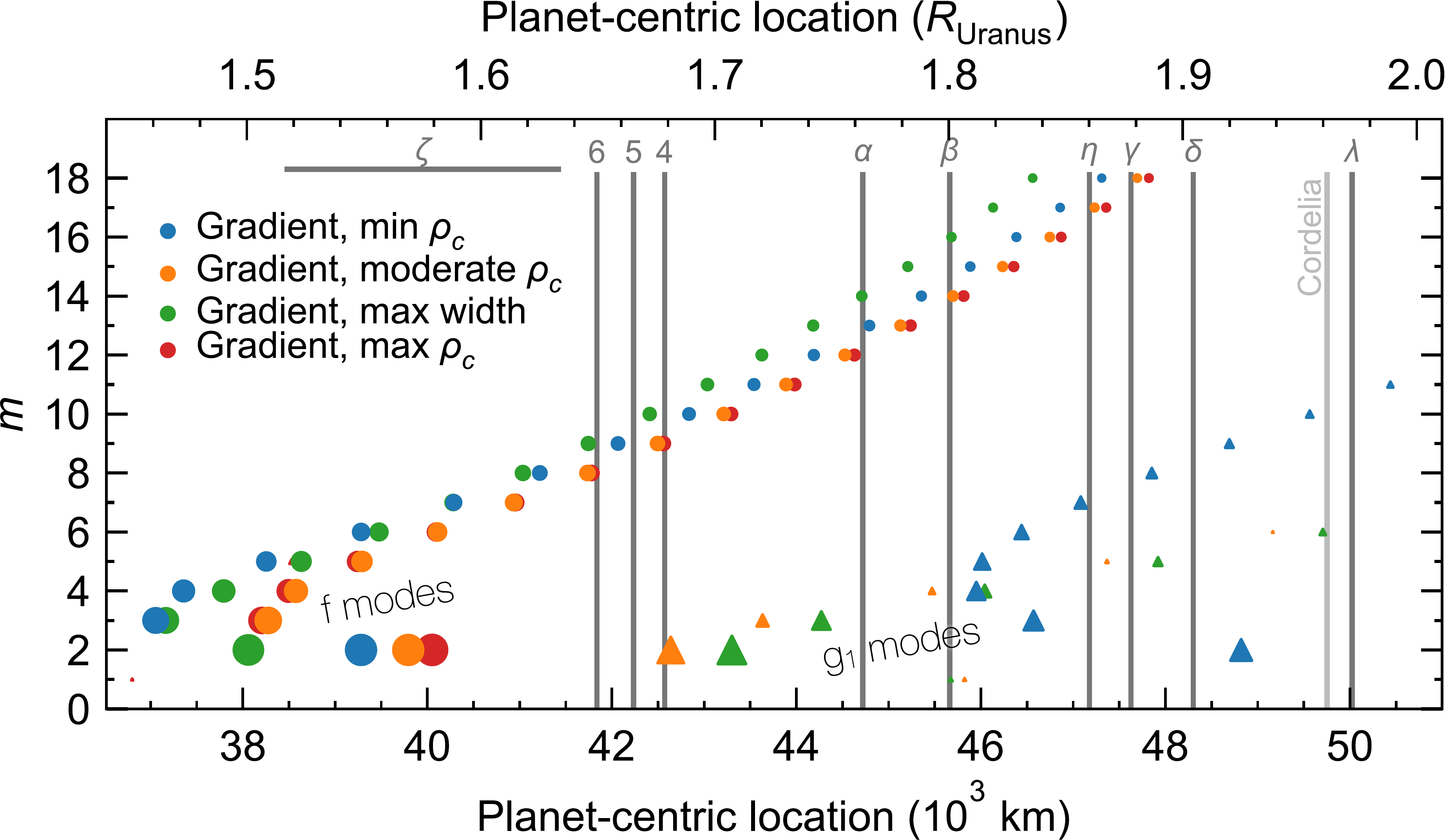}
        \caption{
            \label{fig.ring_olrs_gradient}
            As in Figure~\ref{fig.ring_olrs_interface}, but for gradient models. 
            {Circles denote the f modes.}
            {The modes concentrated at low $m$ and orbital distances $\gtrsim42\times10^3$~km are $n=1$ g modes (triangles). 
            Higher radial order ($n>1$) g modes are expected to produce weaker potentials; these and any g$_1$ modes with $\left|\Phi^\prime(r=R)\right|<10^{-5}\, GM/R$ are omitted here.
            }
        }
    \end{center}
\end{figure*}

Figures~\ref{fig.ring_olrs_interface} and \ref{fig.ring_olrs_gradient} show OLR locations for Uranus modes in our representative set of interface and gradient models respectively, serving as a general guide to the observability of Uranus normal modes in the rings. To facilitate comparison between the models, only sectoral ($\ell=m$) modes are shown. At $\ell>m$ the spectra are dominated by more superficial f modes and hence the become more similar from one interior model to the next, similar to Figure 3 of \citealt{2022PSJ.....3..194A}. Sectoral modes thus likely represent the best opportunity for discriminating interior structure with the detection of a small number of modes. {Nonetheless the existence of $\ell>m$ (non-sectoral) modes, which are omitted from Figures~\ref{fig.ring_olrs_interface} and~\ref{fig.ring_olrs_gradient}, increases the odds of finding planet-associated modes in the rings.}

{The possibility of degeneracy as a function of $\ell-m$ should be addressed. Supposing a resonance is observed and its pattern speed and $m$ value measured, how can it be ascribed to a unique planet mode without direct knowledge of $\ell$? For f modes the interpretation is typically clear because their frequencies are well separated as a function of $\ell$, per the $\omega_f\sim\ell^{1/2}$ scaling of Equation~\ref{eq.f_mode_frequency}. Rotational splitting (Equation~\ref{eq.inertial_frame_frequency}) and planetary structure also affect these frequencies, but rarely to the point of inducing confusion between successive $\ell-m$ values for a given observation (see Figure 3 in either \citealt{2019ApJ...871....1M} or \citealt{2022PSJ.....3..194A}).} 

{For non-f modes like g or interface modes, the dependence of mode frequencies on the compositional or thermodynamic structure introduces degeneracy between interior structure and the unknown $\ell$ value of the mode responsible. In these cases some guidance is provided by the tendency of higher $\ell$ modes to have shorter radial wavelengths, leading them to evanesce more rapidly between their wave propagation zones and the planetary surface. Furthermore, higher $\ell$ components of the perturbed gravitational potential decay more rapidly with radius in the vacuum exterior to the planet. Hence, all else being equal, the lowest allowable angular degree is a priori the most likely to generate a detectable perturbation in the gravity field outside the planet. It remains possible however that the processes responsible for mode excitation and damping could favor certain wavelengths (or $\ell$ values), confounding these simple expectations.}

Among the five interior models considered by \cite{2022PSJ.....3..194A}, only one model had the $_2^2$f mode OLR near a narrow ring: their shallow gradient model and the 5 ring {(see their Figure 3)}. The remainder had faster pattern speeds for $_2^2$f, putting the OLR within the diffuse $\zeta$ ring closer to Uranus, far from any narrow ring. Our models fall exclusively into this latter category: $_2^2$f does not fall among the narrow rings, except in cases with a perfectly rigid core\footnote{{Note that in our models, composition gradients as shallow as A'Hearn et al.'s ``shallow'' model overestimate $\rhoone$ by a factor of several and are hence ruled out. It is possible that a more physically realistic equation of state would allow a larger range of OLR locations with more potential ring overlaps, even in a purely fluid Uranus. }}. Hence if any Uranus mode $m=2$ OLR is observed among the narrow rings, we would interpret it as either a non f mode, or as an f mode with a frequency substantially modified by the presence of a frozen core. We pursue both of these possibilities. 

The possibility of a frozen core in Uranus (e.g., \citealt{2021PSJ.....2..222S}) that does not participate in oscillations close to the f mode frequencies opens up an {wide} range of possible mode spectra. This is because an inert core effectively truncates the f mode cavity from the bottom, dramatically altering the spectrum of f mode frequencies in a manner that is sensitive to the location of the core boundary. 

For deep core boundaries (e.g., moderate to max $\rho_c$; orange and green points in Figure~\ref{fig.ring_olrs_interface}) the f mode spectrum is weakly if at all sensitive to the core state being rigid versus fluid. However for a shallower core boundary $\approx0.7\,\ru$ (min $\rho_c$; blue points in Figure~\ref{fig.ring_olrs_interface}), a frozen core impinges on the region hosting f modes, radically reducing the f mode frequencies, moving for instance the $\ell=m=2$ f mode OLR from $40.0\times10^3~{\rm km}=1.58~\ru$ to $50.4\times10^3~{\rm km}=1.99~\ru$. Toward higher $\ell=m$ the effect is gradually diminished, but intermediate $\ell=m=3-7$ modes in this model notably have their OLRs moved from a region devoid of narrow rings into near resonances with the 6, 5, 4, and $\alpha$ rings. Therefore, if Uranus's interior structure resembles our interface model, then the detection of a single low to intermediate $m$ mode among the narrow rings would be a powerful discriminant of the core state and the radial location of the core boundary.

However, degeneracies follow from the nonuniqueness of the model. The equally valid gradient models depicted in Figure~\ref{fig.ring_olrs_gradient} have f mode OLR locations broadly similar to the interface models, but also host g modes that can generate structure in the rings. Three of the four models have g$_1$ modes (i.e., $n=1$ g modes) at frequencies lower than the f mode, with OLRs dotting the landscape of the narrow main rings. The fourth (red) model is that with the strongest stable stratification at the core transition, giving rise to g$_1$ mode frequencies {comparable to or in excess of} the f mode frequencies. In this case the $\ell=m=2,3$ OLRs are well interior to the narrow rings{, falling outside the range of Figure~\ref{fig.ring_olrs_gradient}}. Indeed, the data permit a continuum of intermediate interior models that can yield g$_1$ modes throughout the main rings. Just as with the interface models discussed above, the detection of a single low to intermediate $m$ g mode among the narrow rings could decisively rule out most of the models, at least within the confines of a specific parameterization for the interior. {Note that the g modes in the model with the deepest composition gradient (maximum central density; red points in Figure~\ref{fig.ring_olrs_gradient}) yield substantially weaker external potentials as they evanesce over the intervening convective regions. Hence, even putting their frequencies aside, the detectability of g modes in the rings hinges on the stable stratification not being confined too close to the planetary center.}

{
    Before we proceed, we note that prograde Uranus modes can also generate ILRs among the rings. 
    Unlike OLRs, these inner resonances require pattern speeds \textit{slower} than the ring mean motion, leading to a preference for f modes of high $m$, or g or interface modes with low to moderate $m$.
    Indeed, \cite{2021Icar..37014660F} discovered matched OLR-ILR pairs in Saturn's C ring, each generated by a single high-$m$ f mode of Saturn.
    For the 7 models presented in Figures~\ref{fig.ring_olrs_interface}-\ref{fig.ring_olrs_gradient}, we find $\ell=m=14$--$17$ f mode ILRs in the vicinity of the 6, 5, and 4 rings. 
    Higher $m$ f modes could yield ILRs in more distant rings.
    For interface models, the interface modes lead to a wide variety of ILRs throughout the rings, especially for shallower interfaces.
    For gradient models, $g_1$ modes with $\ell=m=7$--$9$ have ILRs near the 6, 5, and 4 rings, with their higher $m$ counterparts falling throughout the more distant rings.
    In what follows we focus on the OLRs, where the overlaps with ring orbits are more common, but stress that ILRs are not unexpected and are generally as useful as their OLR counterparts.
    The same goes for inner and outer vertical resonances (IVRs/OVRs), which can arise from vertical perturbations by north-south asymmetric (odd $\ell-m$) modes; see \cite{2022PSJ.....3..194A}.
}

\subsection{Fits to hypothetical ring resonances}\label{sec.rings.fits}
Here we suppose that high-resolution imaging reveals a forced $m=2$ distortion in the 6 ring, betraying the influence of a nearby {Lindblad resonance} with an $m=2$ driving potential. In lieu of any plausible interpretation in terms of mean motion resonances with known satellites, a Uranus $\ell=m=2$ normal mode OLR may be the best explanation. The 6 ring's mean semimajor axis \citep{2018prs..book...93N}\footnote{See \cite{2024Icar..41115957F} for refined ring orbital elements and width-radius relations; the differences are not significant for this purpose.} implies that an $m=2$ OLR has a pattern speed $\omp=2089.3~\degd$. Here we explicitly incorporate this frequency as an added constraint on the interior model, augmenting our usual likelihood (see Section~\ref{sec.methods}) with a Gaussian term with standard deviation equal to $10^{-3}$ times the centroid pattern speed. This amounts to an OLR location $1\sigma$ uncertainty of order 10 km{. This uncertainty is chosen to encompass a range of resonance locations that could reasonably influence a narrow ring:} for comparison, satellite-driven ring modes are {typically} evident {within a few km of} their associated resonances \citep{2024Icar..41115957F}. 

\begin{figure}
    \begin{center}
        \includegraphics[width=\columnwidth]{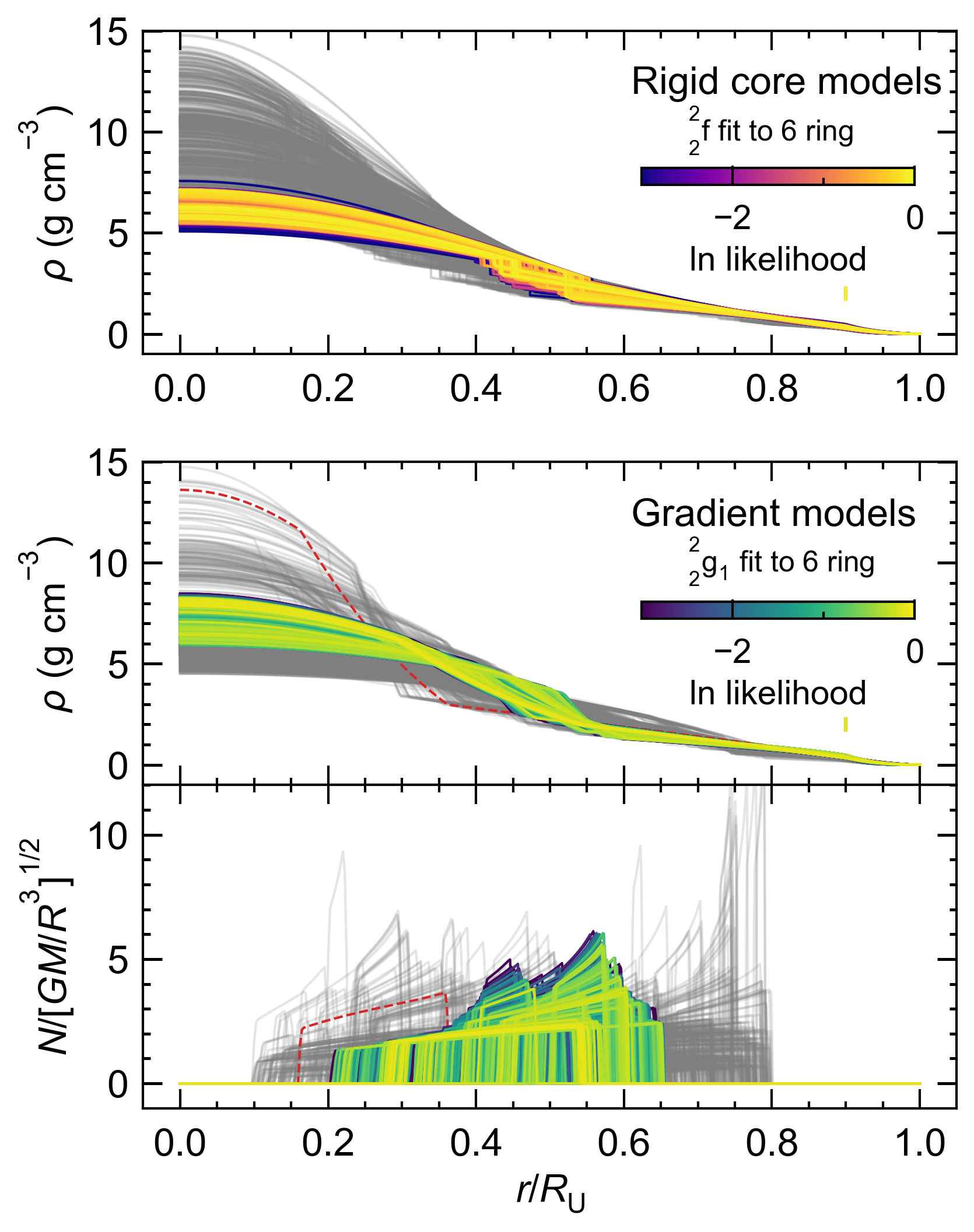}
        \caption{
            \label{fig.6ring_fits_interiors}
            Interior profiles constrained by a hypothetical ring seismology constraint (curves with color mapped to likelihood) compared to those of their parent sample unconstrained by seismology (grey curves).
            The top panel shows density profiles $\rho(r)$ for interface models with a rigid (non-oscillating) core; the middle and bottom panels show $\rho(r)$ and $N(r)$ respectively for models with a composition gradient.
            Models in grey are constrained solely by $J_2$, $J_4$, and $\rhoone$; color coded models additionally fit a single $m=2$ ring seismology constraint as described in the text.
            Only a randomly chosen subset of the models---1,024 per sample---are shown.
            The red dashed curves in the gradient case (lower panels) show the best profile yielded by an alternative sample that identifies the same resonance with the higher order g mode $_2^2$g$_2$ mode instead.
            }
        \end{center}
    \end{figure}
    
\begin{figure*}
    \begin{center}
        \includegraphics[width=0.7\textwidth]{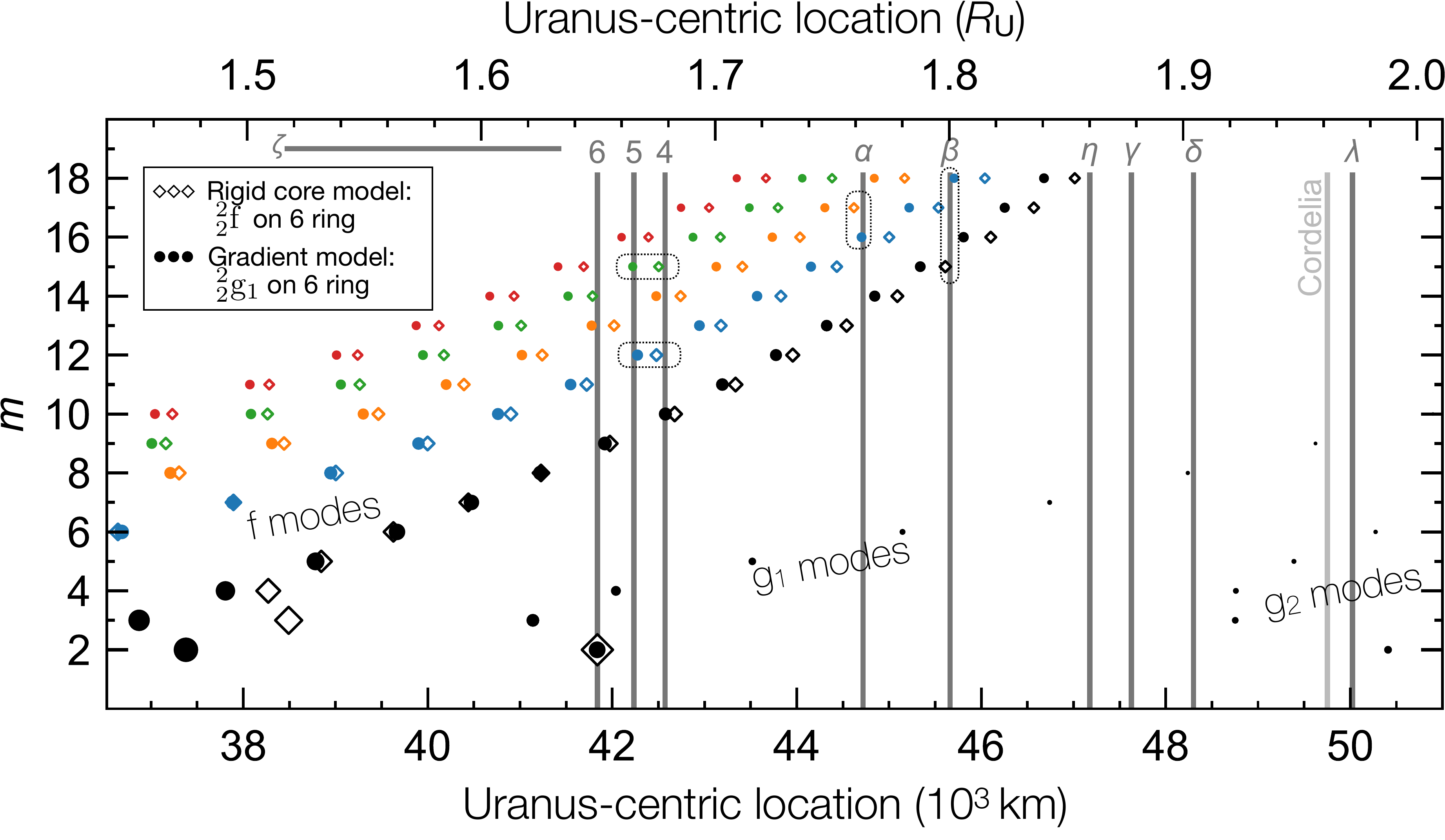}
        \caption{
            \label{fig.6ring_fits_olrs}
            OLR locations amid the rings for the best models obtained directly fitting either the $_2^2$f mode (in the rigid core model, open diamonds) or the $_2^2$g$_1$ mode (in the gradient model, filled circles) to an $m=2$ OLR near the 6 ring.
            Modes with $\ell=m$ (black) are shown, as are modes with $\ell-m=2$ (blue), 4 (orange), 6 (green), and 8 (red). 
            This latitudinal wavenumber $\ell-m$ generally increases toward the top left of the diagram.
            The dashed boxes highlight additional associations that would discriminate between the two models shown here (see text).
        }
    \end{center}
\end{figure*}
    
The interior profiles of the resulting distributions are shown in Figure~\ref{fig.6ring_fits_interiors}, where the models benefiting from the single ring seismology constraint (colormaps) are superimposed on the corresponding sample fit to $(J_2,J_4,\rhoone)$ alone (grey). Clearly even a single OLR detection near a narrow ring can greatly restrict the landscape of permissible interior models. 

By comparing the results of these two interior model parameterizations applied to the same pair of observables $(m,\omp)$, the figure also highlights the ambiguity in interpeting a single mode: the solutions change depending on whether we identify it with an f mode or a $g_1$ mode. Notably, these two scenarios happen to yield similar density profiles in Uranus: the rigid core models produce $\rcore=0.486\pm0.040\,\ru$ (mean and standard deviation) and the gradient models produce equivalent core radii $(r_o+r_i)/2=0.473\pm0.037\,\ru$. Still other interpretations of the same data could produce more radically different solutions for Uranus's interior structure, but based on the relatively large gravitational potentials associated with f and $n=1$ g modes, we consider the two scenarios shown here to be the most likely. {In any case, a low-degree seismology constraint has the potential to constrain the core radius to within a few percent, a striking improvement over the totally unconstrained core boundary in the absence of seismology (see Appendix~\ref{app.parameter_estimation}).}

Figure~\ref{fig.6ring_fits_olrs} presents the full spectrum of OLRs in the rings for the best fitting model in each case. Looking beyond the $m=2$ resonances, the spectra differ substantially from one another, suggesting that the measurement of a second mode frequency at higher $m$ could eliminate the degeneracy between the two scenarios. For example, the gradient model produces a $_{14}^{12}$f mode OLR just exterior, and a $_{21}^{15}$f mode OLR just interior, to the 5 ring. The detection of a $m=12$ or $m=15$ forcing on the 5 ring would strongly disfavor the rigid core model, which locates these two OLRs closer to the 4 ring. Similarly, the most likely seismic signature on the $\alpha$ ring is predicted to be $m=17$ (by virtue of $_{21}^{17}$f) in the rigid core model, but $m=16$ (excited by $_{18}^{16}$f) in the gradient model. On the $\beta$ ring one would look to distinguish between an $m=15$ signal ($_{15}^{15}$f, rigid core) and an $m=18$ signal ($_{20}^{18}$f, gradient model).

Inspection of Figure~\ref{fig.6ring_fits_olrs} invites a third interpretation, ascribing the same $m=2$ resonance at the 6 ring to the $_2^2$g$_2$ resonance in the gradient model{, which appears a distant $50.4\times10^3$ km from Uranus's center for the model in the diagram. Moving the $n=2$ g mode Lindblad resonances closer to Uranus} is possible if the g mode spectrum is pushed to higher frequencies by increasing the typical value of $N$ within the gradient region. This requires some combination of reducing the gradient's radial extent $r_o-r_i$ and enhancing its density contrast $\agrad$. Indeed, we find {that} a sample that fits the $_2^2$g$_2$ pattern speed to a 6 ring OLR strongly prefers compact cores $(r_o+r_i)/2=0.265\pm0.032\,\ru$ with high central densities $\rho_c=13.8\pm0.7\,{\rm g\ cm}^{-3}$. The best single model is shown in Figure~\ref{fig.6ring_fits_interiors} (bottom panel, dashed red curve). Nonetheless we reiterate that f and $n=1$ g modes are much more likely a priori to generate observable signatures in the rings because of their larger intrinsic gravity perturbations. 
{For comparison, assuming all modes in the gradient model of Figure~\ref{fig.6ring_fits_olrs} possess equal energy, the surface potential perturbation of $_2^2{\rm g}_2$ is approximately 1/5 that of $_2^2{\rm g}_1$ and $1/10$ that of $_2^2{\rm f}$.}

\begin{figure*}
    \begin{center}
        \includegraphics[width=0.61\textwidth]{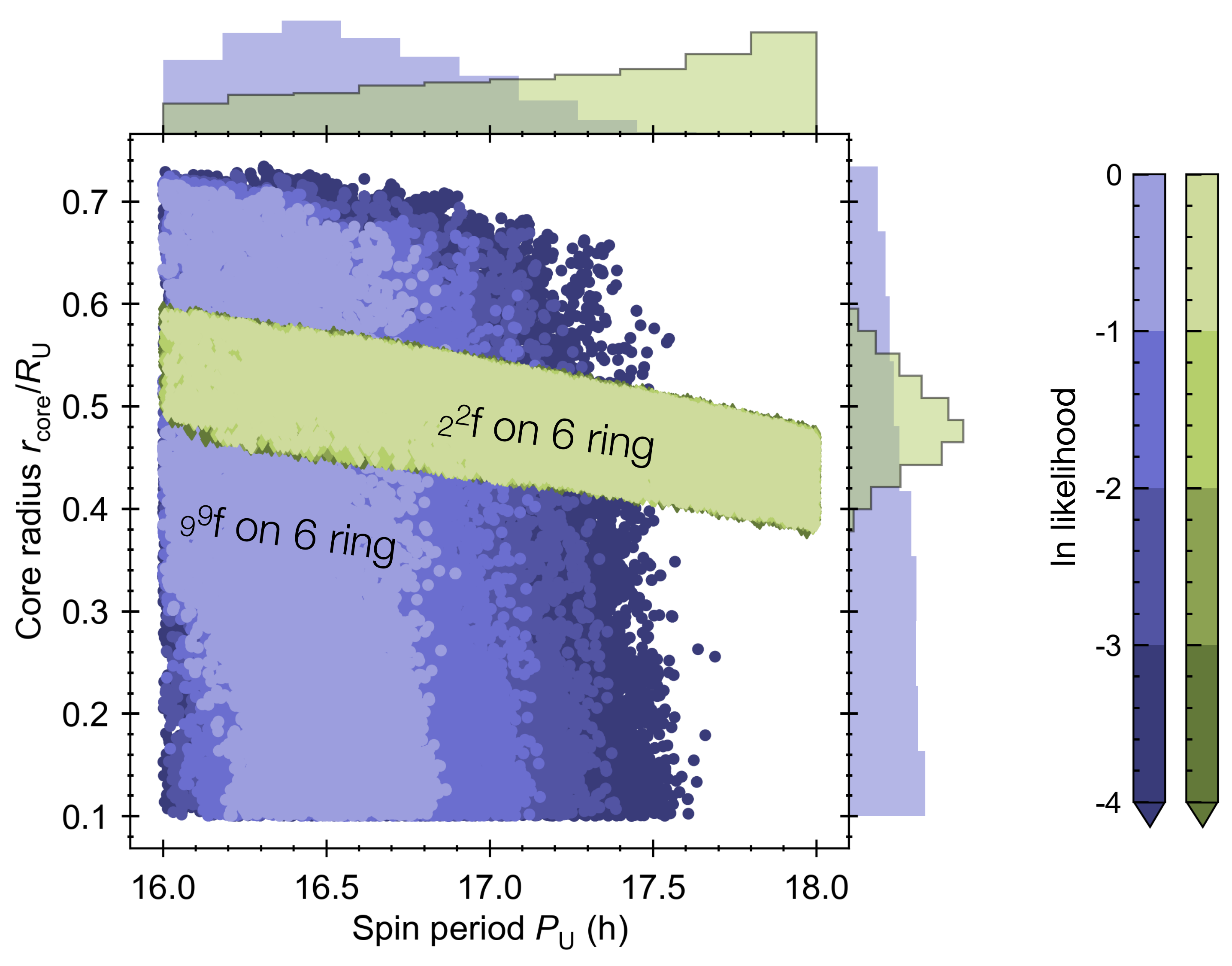}
        \caption{
            \label{fig.6ring_99f_fit_scatter}
            {
            Core radius versus bulk rotation period in a sample fitting a high degree mode ($_9^9$f) to the 6 ring (purple colormap), compared to that fitting a low degree mode ($_2^2$f) to the same ring (green colormap; this is the same sample as the rigid core sample seen in Figure~\ref{fig.6ring_fits_interiors} and \ref{fig.6ring_fits_olrs}). 
            The low-degree constraint would constrain Uranus's core radius, but not its rotation. 
            The high-degree f mode is more superficial and would be insensitive to core properties, but would constrain Uranus's rotation period.
            }
            }
        \end{center}
    \end{figure*}

{Admittedly, among the modes that may resonate with the rings, the deeply penetrating $_2^2$f mode may be one of the most optimistic scenarios for constraining Uranus's core structure. 
We perform an additional experiment that instead fits the high angular degree mode $_9^9$f to the 6 ring, again taking Uranus's core as perfectly rigid to allow a direct comparison with the rigid core model seen in Figures~\ref{fig.6ring_fits_interiors}-\ref{fig.6ring_fits_olrs}. 
Figure~\ref{fig.6ring_99f_fit_scatter} gives the posterior probability distributions of $\rcore$ and $\pu$, with marginalized 1D histograms of each, comparing the sample with $_2^2$f fit to data to the one with $_9^9$f fit to data.
The $_9^9$f sample indeed leaves the core properties poorly constrained, a reflection of this more superficial mode's small amplitudes in the core. 
However, in this case the single seismology constraint leads to a strong preference for bulk rotation periods faster than 17 h, a powerful result in its own right.
Quantitatively, the $_9^9$f sample favors $\pu>17\,{\rm h}$ at $87\%$ confidence, compared to the more rotation-ambivalent $_2^2$f sample which favors $\pu>17\,{\rm h}$ at 63\% confidence.
The greater sensitivity of the high-degree modes follows from modes of higher $m$ attaining larger frequency shifts as a result of Uranus's rotation via Equation~\ref{eq.inertial_frame_frequency}.
The detection of a high-$\ell$, preferably $m=\ell$ mode, thus offers an opportunity to constrain Uranus's unknown internal rotation, breaking a degeneracy that measurements of the zonal gravity moments alone cannot resolve.}

{
We note that the notional pattern speed uncertainty $\sigma_{\omp}/\omp=10^{-3}$ ($\sim10$ km in OLR location) we have adopted is likely to be overly pessimistic: from Voyager and ground-based occultations, \cite{2024Icar..41115957F} have estimated (typically satellite associated) Lindblad resonance locations with sub-km precision.
Hence, a ring seismology based estimate for Uranus's rotation may be substantially more precise than the purple distribution displayed in Figure~\ref{fig.6ring_99f_fit_scatter}.
}

\subsection{Interface modes and avoided crossings in the rings}\label{sec.rings.avoided}
We have demonstrated that a{n extended} rigid core can shift low $m$ f mode OLRs into the narrow rings. The interface models with \emph{fluid} cores also can produce additional (non f mode) structure in the rings by virtue of the interface modes {hosted at their core boundary}. These are not overtly obvious in the three models depicted in Figure~\ref{fig.ring_olrs_interface}, but a close inspection of the $r\lesssim 1.5\,\ru$ region reveals a faint sequence of $m=5,6,\ldots$ modes in the moderate $\rho_c$ fluid core model. In fact, for core boundaries within an appropriate range, interface modes can resonate throughout the narrow rings.

\begin{figure*}
    \begin{center}
        \includegraphics[width=0.8\textwidth]{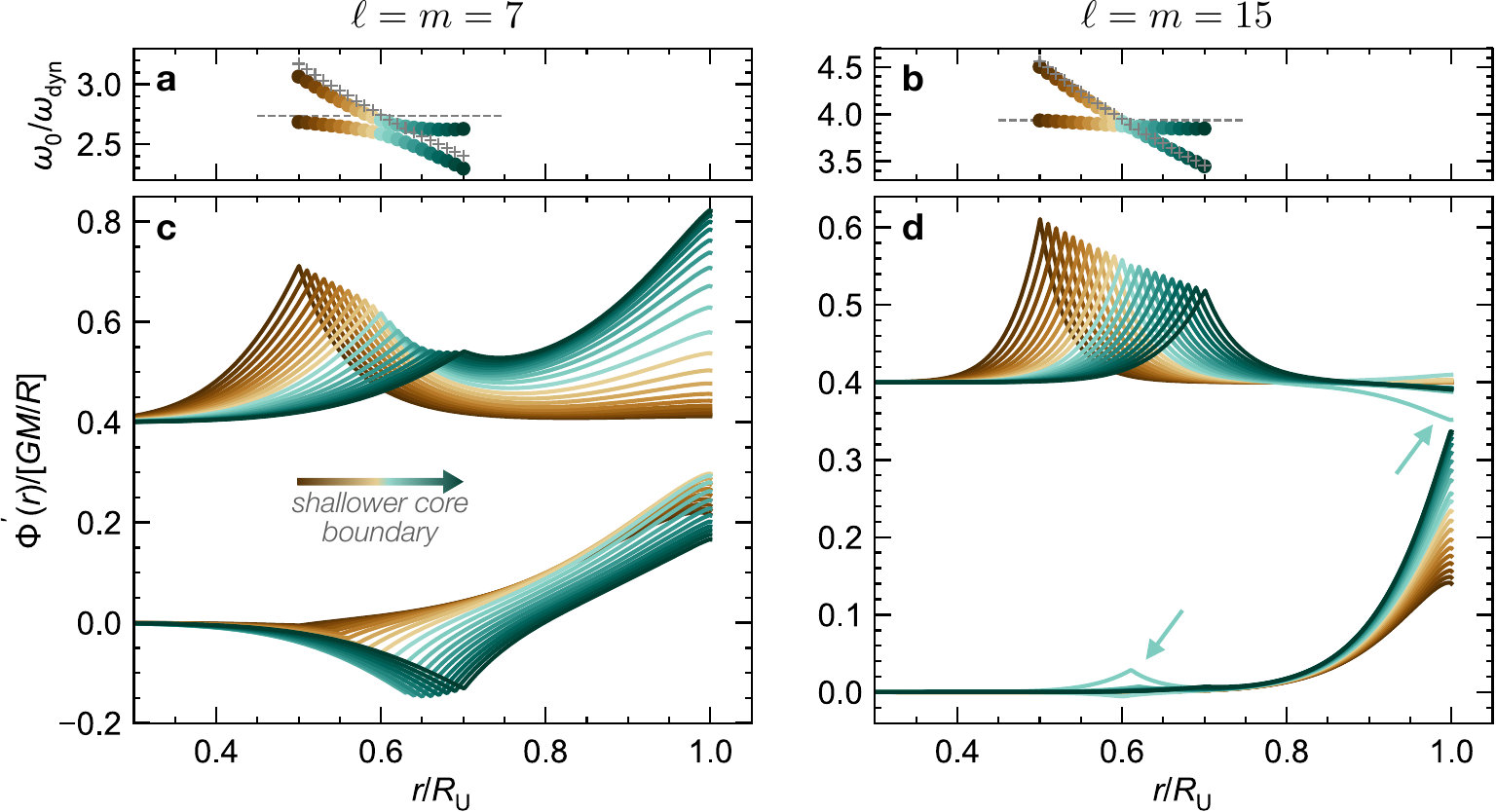}
        \caption{
            \label{fig.fi_avoided_crossings_eigs}
            {Frequencies and e}igenfunctions of the gravitational potential perturbation $\Phi^\prime$ for interface and f modes with similar frequencies, in a grid of models varying the core boundary between $0.5$ {(dark brown)} and $0.7\ \ru$ {(dark green)}.
            {\textbf a}, {\textbf c}: $\ell=m=7$ modes with strong eigenfunction overlap giving rise to mixed interface/f mode character and an avoided frequency crossing.
            {The higher frequency mode (upper $\Phi^\prime$ eigenfunction, offset vertically by $+0.4$) transitions from interface-dominated to f-dominated as the core boundary is increased.}
            {\textbf b}, {\textbf d}: $\ell=m=15$ modes with little to no eigenfunction overlap; in this case the interface modes (upper eigenfunctions) remain essentially decoupled from the f modes (lower) and are unlikely to be observed.
            Arrows highlight the model yielding the smallest frequency separation, inducing weak mode mixing.
            {The frequency axes ({\textbf a}, {\textbf b}) show the centroid ($m=0$) mode frequencies as a function of core radius.}
            Cross symbols represent the analytic approximation to the frequency of the interface mode (Equation~\ref{eq.interface_mode_frequency}) and the dashed horizontal line gives the approximate frequency of the f mode (Equation~\ref{eq.f_mode_frequency}).
        }
    \end{center}
\end{figure*}

\begin{figure}
    \begin{center}
        \includegraphics[width=\columnwidth]{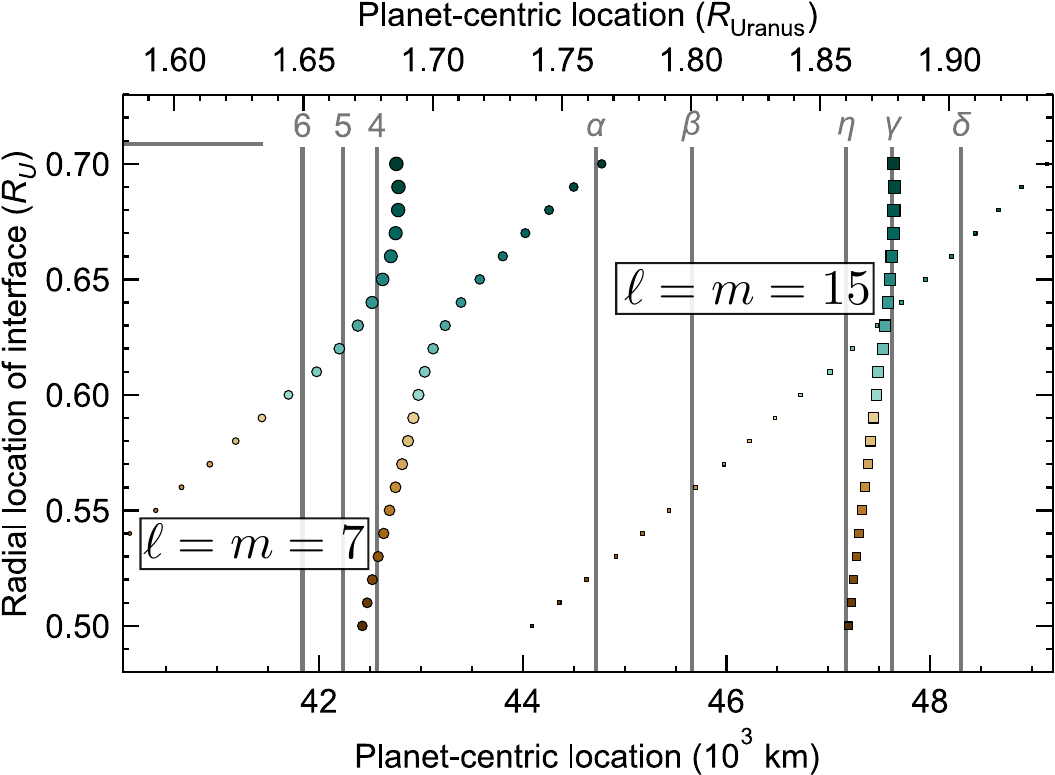}
        \caption{
            \label{fig.fi_avoided_crossings_olrs}
            Outer Lindblad resonance (OLR) locations for the modes highlighted in Figure~\ref{fig.fi_avoided_crossings_eigs}.
            Point size scales with strength of the gravity perturbation at Uranus's surface.
            At $\ell=7$ {(circles)} the deeply penetrating f modes interact strongly with modes trapped on the core interface, giving rise to mixed modes that could be observed as a pair of distinct OLRs with the same $m$. 
            At $\ell=15$ {(squares)} the f modes are trapped so closely to the surface that f and interface modes are effectively decoupled.
        }
    \end{center}
\end{figure}

Interface modes are most likely to be observed when their frequencies are similar to the f modes frequencies, leading to degenerate mode mixing and producing mixed interface/f modes. This phenomenon is similar to the avoided crossings inferred to take place between f and g mode in Saturn based on ring seismology data \citep{2014Icar..242..283F,2021NatAs...5.1103M} or p and g modes in post-main sequence stars \citep{1975PASJ...27..237O,1977A&A....58...41A}. Based on the approximations for f and interface mode frequencies given by Equations~\ref{eq.f_mode_frequency} and \ref{eq.interface_mode_frequency}, the two are equal when
\begin{equation}
    \label{eq.interface_f_mode_crossing}
    \frac{\langle\rho\rangle_i}{\langle\rho\rangle}=\left(\frac{1+\ajump}{1-\ajump}\right)
\end{equation}
where $\langle\rho\rangle_i$ denotes the mean density of the core and $\langle\rho\rangle$ that of Uranus. Notably, this condition is independent of $\ell$, meaning that a model close to satisfying Equation~\ref{eq.interface_f_mode_crossing} {will have overlapping} f and interface mode {frequencies} for all $\ell$. {But as we will show, mode mixing is only significant in cases with sufficient eigenfunction overlap inside Uranus, typically limiting f/interface mixtures to low to moderate angular degrees.}

Figure~\ref{fig.fi_avoided_crossings_eigs} traces the gravitational potential perturbation eigenfunctions for the interface/f mode pair over a sequence of gravity-constrained models with $\ajump=2$ and $\rcore=0.5-0.7\,\ru$. Even relatively large fractional frequency separations can yield strong mode mixing, if the eigenfunctions of the modes in question have sufficient overlap. This is the case for the $\ell=m=7$ modes, where the frequencies of isolated interface modes (colorful crosses; Equation~\ref{eq.interface_mode_frequency}) would sweep through the frequency of the isolate{d} f mode (horizontal dashed line; Equation~\ref{eq.f_mode_frequency}), the true frequencies repel one another as the eigenfunctions attain a mixed character. The interaction with the f mode tends to amplify the gravity peturbation of the interface modes, producing more potentially observable signatures in the rings. In contrast, at higher angular degree $\ell=m=15$ the interface and f mode eigenfunctions are so well confined to their respective regions of propagation that even a close frequency crossing can induce only weak mixing. As a result, the frequency sequences can cross essentially unimpeded and the interface modes remain trapped within the planet, their surface gravity perturbations too small to be likely to be observed.

Figure~\ref{fig.fi_avoided_crossings_olrs} summarizes the signatures that interface/f mode crossings may produce in the rings. The interface mode has little effect on the $\ell=m=15$ spectrum, where the f mode remains the only mode likely to be observed, in this case through its resonant influence on the $\eta$ or $\gamma$ rings. In the $\ell=m=7$ spectrum, however, each model can produce a pair of OLRs with observable amplitudes. One one hand, this again exposes the degeneracy in detecting a single Uranus mode: if observations reveal an $m=7$ potential forcing the 4 ring, this model would be unable to discern between a core boundary at $0.53\,\ru$ and one at $0.65\,\ru$. On the other hand, the fact that mode mixtures can produce well-separated doublets of OLRs with the same $m$ value raises the possibility that two resonances influencing separate rings could be connected to a single feature of Uranus's interior. For instance, the model with a core boundary at $0.70\,\ru$ has an f-dominated mixed mode OLR near the 4 ring and an interface-dominated mixed mode OLR near the $\alpha$ ring. If both rings showed evidence of $m=7$ forcing this could {present} powerful support for a core boundary near $0.7\,\ru$.

Different models in our samples manifest f/interface mode crossings with different observable OLR pairs, for example coinciding with the 6 and 4 ring. Hence the picture in Figure~\ref{fig.fi_avoided_crossings_olrs} is only one possibility, but serves to emphasize that a pair of observed resonances with the same $m$ value may follow from mixed modes in the interior, and our models can be used to select between the possible interpretations. 

Finally, we note that very similar mode mixing behavior can take place in our gradient models, between f and low order g modes. The two have similar frequencies when typical values of $N$ exceed approximately $(\ell)^{1/2}\omega_{\rm dyn}$, where $\ell$ is at least 2 for the f modes. Ring seismology suggests that this is the case in Saturn (e.g., \citealt{2014Icar..242..283F,2021NatAs...5.1103M}) and it is also the case for a subset of our Uranus gradient models. We do not delve into f/g mode mixing in detail but its implications for the rings are similar to the f/interface mode mixing above and of course to the well-studied f/g mixed modes evident in Saturn's rings (see also \citealt{2022MNRAS.516..358D}). This possibility will need to be entertained as new Uranus ring data are analyzed.

\subsection{Are ring constraints valuable despite improved zonal gravity?}\label{sec.rings.fits_improved_gravity}
{Radio tracking of a}n orbiter {could} yield orders of magnitude improvement in the precision on $J_2$ and $J_4$ and offer the first measurements of the higher order moments $J_{6+}$ {\citep{2024PSJ.....5..116P}}. 
Even in light of these improvements, we find that the measurement of a single normal mode frequency via the rings would contribute a critical independent constraint on the density and extent of Uranus's core.
Here we fit interface models using $\sigma_{J_2}=0.0073\times10^{-6}$, $\sigma_{J_4}=0.9935\times10^{-6}$ reflecting conservative estimates expected from gravity orbits by a Uranus orbiter (\citealt{2024PSJ.....5..116P}; see Section~\ref{sec.methods.models.constraints}). 

We put aside higher order moments $J_{6+}$, recognizing that (a) their centroid values are unknown, (b) their sensitivity to the deep interior is limited compared to their low order counterparts, and (c) their wind-induced contributions might be comparable to those from the rigidly rotating background structure. 
{On this last point we note that our baseline sample of gradient models yields $J_6^{\rm rigid}=(0.62\pm0.08)$ ppm and $J_6^{\rm winds}=(-0.25\pm0.16)$ ppm, for a total $J_6=(0.38\pm0.11)$ ppm. The higher degree $J_n$ will have even larger fractional contributions from the winds, making the values of $J_{6+}$ ambiguous with regard to internal structure.
This situation with $J_6$ being the `pivotal' even-degree moment between the structure-dominated and wind-dominated moments is reminiscent of Saturn. 
One might hope that Uranus's evidently shallower winds would improve the ability of the $J_{2n}$ to probe interior structure, but this is countered by the diminished rigid body moments that follow from Uranus's much weaker oblateness.
}

\begin{figure}
    \begin{center}
        \includegraphics[width=\columnwidth]{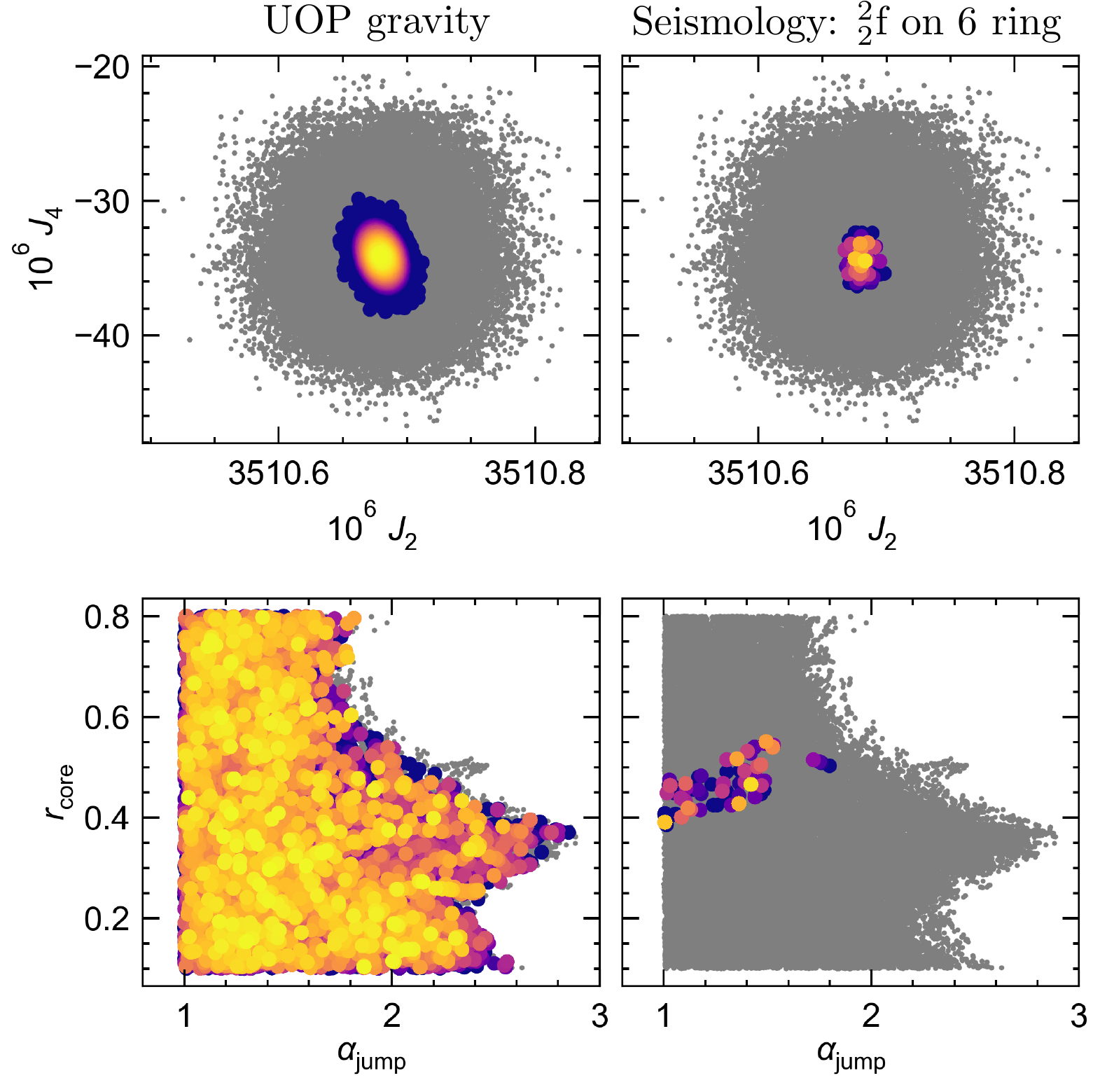}
        \caption{
            \label{fig.rejection_resampling_fit_mode}
            Gravity moments (upper panels) and core properties (lower panels; core density enhancement versus core radius) for Uranus models in 3 scenarios constrained by progressively more information.
            Grey samples in the background are the permissive starting sample loosely constrained to $J_2$ and $J_4$.
            The ``UOP gravity'' distribution (left panels, colormap) benefits from more precise gravity moments from a Uranus orbiter as estimated by \cite{2024PSJ.....5..116P} (see main text for details).
            The seismology sample (right panels, colormap) folds in a hypothetical seismic constraint in the form of a resonance between the $_2^2{\rm f}$ mode and the 6 ring.
            {Colors correspond to log likelihood using the same scale as Figure~\ref{fig.6ring_fits_interiors} (top).}
        }
    \end{center}
\end{figure}

These precise gravity moments proved challenging for our normal sampling process{, leading us to develop a different procedure based on rejection sampling. 
The methodology is described in Appendix~\ref{app.rejection_sampling}; here we focus on the results. 
The nature of our algorithm leads to a relatively small number ($1,207$) of models in the final seismology-informed sample, from which we do not intend to draw robust quantitative statistical conclusions.
Nevertheless, }this procedure is sufficient to show that the single ring seismology data point powerfully restricts the range of allowed interior models, even when gravity moments are already known to high precision.
Figure~\ref{fig.rejection_resampling_fit_mode} compares the permissive sample, the sample constrained by orbiter gravity, and the sample constrained jointly by orbiter gravity and seismology.
Despite the much smaller scatter in $J_2-J_4$ space afforded by the orbiter gravity (upper left panel), the {core} parameters (lower left model) remain highly degenerate, yielding little in the way of constraining power for the deep interior of Uranus. 
In contrast, an association between the $_2^2$f mode and the 6 ring severely restricts the range of allowed models, establishing an upper limit $\ajump<1.8$ to the fractional density contrast across the core boundary and locating the core boundary between $0.39$ and $0.55\,\ru$.

\begin{figure}
    \begin{center}
        \includegraphics[width=\columnwidth]{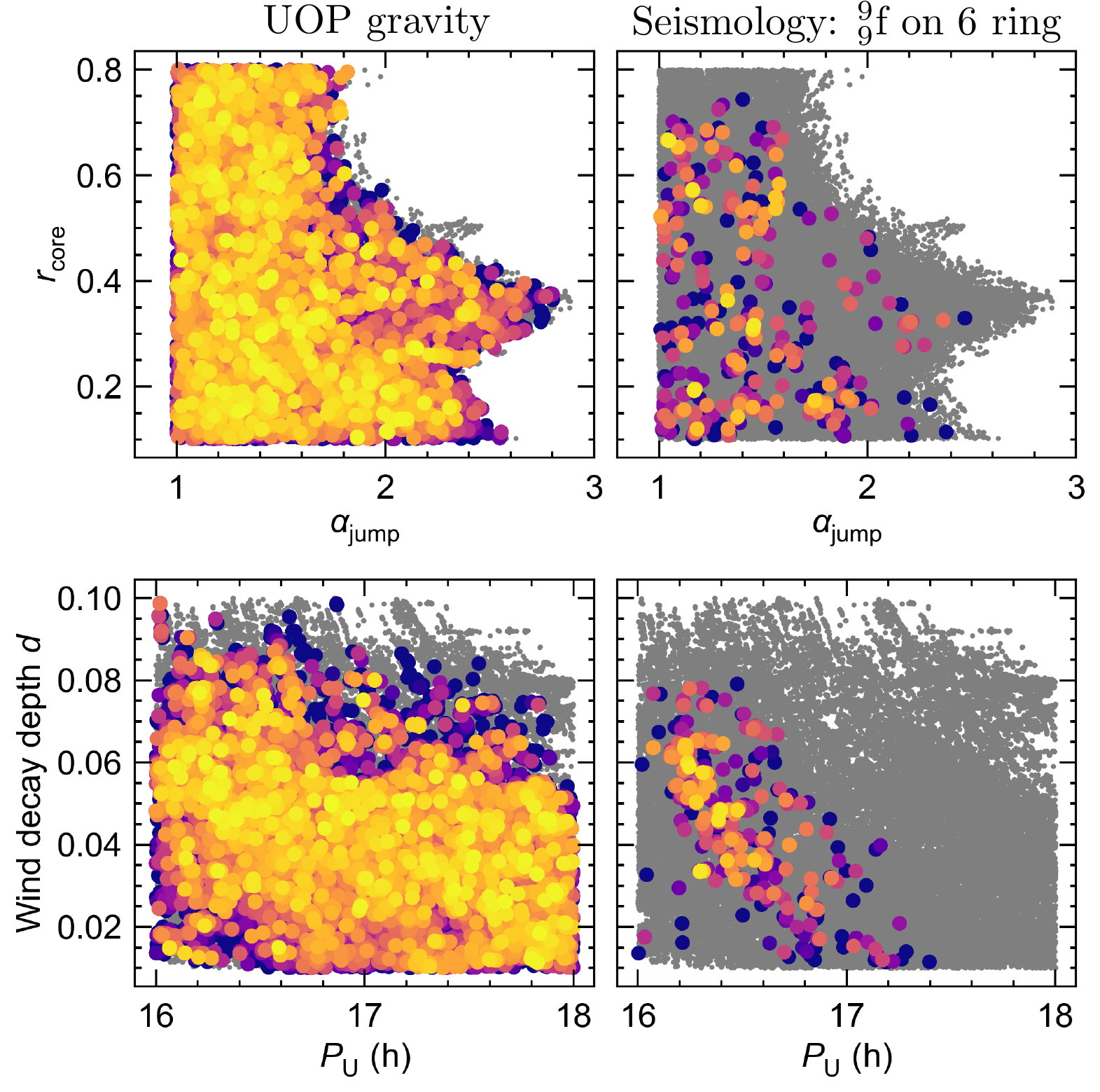}
        \caption{
            \label{fig.rejection_resampling_fit_mode_high_l}
            Similar to Figure~\ref{fig.rejection_resampling_fit_mode}, but here the seismology sample (right panels, colormap) posits a resonance between the 6 ring and the higher degree $\ell=m=9$ f mode.
            Here upper panels show the core density enhancement versus core radius, and lower panels show the wind decay depth versus deep spin period.
            In this scenario the single seismology data point would do little to constrain Uranus's core structure, but would directly constrain Uranus's bulk rotation.
        }
    \end{center}
\end{figure}

{In Section~\ref{sec.rings.fits} we considered the possibility that a high-degree f mode constraint might be detected instead of a low-degree f mode. Here we repeat the exercise of fitting the $_9^9$f mode to the 6 ring, but assuming the stricter UOP zonal gravity constraints. The results are seen in Figure~\ref{fig.rejection_resampling_fit_mode_high_l}. As before, this more superficial mode is unable to constrain core properties, but does strongly disfavor deep spin periods slower than approximately 17 h. A high-degree f mode in the rings could hence be used to break the rotation degeneracy that even a highly precise gravity determination would not.}

We conclude that improved constraints on zonal gravity do not diminish the value of Uranus ring seismology. As is the case at Saturn, gravity science and seismology are most effectively analyzed in tandem \citep{2023PSJ.....4...59M}.

\section{Doppler imaging: p modes}\label{sec.doppler}
Uranian normal modes may also be accessible through Doppler imaging, the technique of creating spatially resolved maps of the radial velocity of the planet's surface. 
These Doppler maps can be decomposed using spherical harmonics, and a time series of the results can be analyzed to generate power spectra associated with each spherical harmonic, from which individual frequencies could be extracted.
(In reality, the data being limited to the visible disk introduces leakage between the spherical harmonics; see \citealt{2002RvMP...74.1073C}).
In principle, this technique allows the characterization of normal modes despite the measured radial velocities being contaminated by radial motion associated with atmospheric dynamics, a rich data set in its own right that was recently charted by \citealt{2024PSJ.....5..100S}. 

This technique has its roots in helioseismology and shows promise for unveiling the normal modes in Jupiter from ground-based observations (e.g., \citealt{2011A&A...531A.104G,2024PSJ.....5..100S,2022FrASS...968452S}).
However, it is extremely unlikely that similar observations from the ground or Earth orbit will be fruitful for Uranus given its faintness and small angular size. 
A Uranus orbiter therefore poses a special opportunity for {sensitive} Doppler imaging seismology of an ice giant.
More detailed practical study of this kind of instrumentation is needed; the challenges of these observations from spacecraft are touched on briefly in Section~\ref{sec.discussion}.
These data are likely most sensitive to p modes, whose higher frequencies imply larger velocity amplitudes than lower frequency modes with similar energies. 
Knowledge of the p mode spectrum would be a powerful means of probing small spatial scale features in Uranus's interior.

\subsection{Échelle diagrams}
\label{sec.doppler.echelle}

\begin{figure*}
    \begin{center}
        \includegraphics[width=\textwidth]{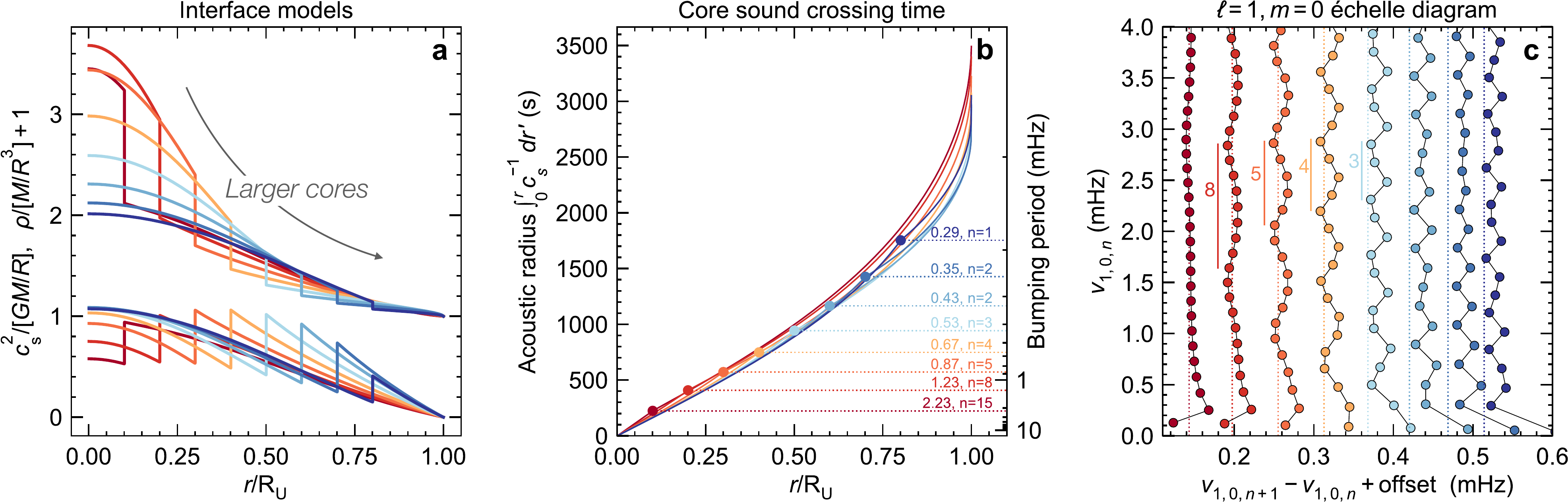}
        \caption{
            \label{fig.echelle}
            Density and sound speed profiles ({\bf a}), profiles of the acoustic radius ({\bf b}) and $\ell=1$, $m=0$ \'echelle diagrams ({\bf c}) for a grid of interface models.
            The density jump amplitude is fixed at $\ajump=2$ and colors map to the assumed core boundary location.
            For legibility the \'echelle diagrams are offset {from the first} by multiples of 0.05 mHz.
            Vertical dotted lines show the asymptotic equal frequency spacing $\Delta\nu$ given by Equation~\ref{eq.delta_nu}{, also offset horizontally}.
            The periodicity imprinted on the frequency spacings is diagnostic of the acoustic radius of the interface (solid circles {and dotted lines} in panel {\bf b}) which translates into the `period' (units mHz, nonlinear right axis of panel {\bf b}) of the bumping signature. Dividing by $\Delta\nu$ provides the equivalent spacing in terms of radial order, which {matches} the bumping in the calculated spectrum (vertical bars in panel {\bf c}).
        }
    \end{center}
\end{figure*}

By way of scientific motivation for an orbiter-borne Doppler imager, here we extend the calculation for our Uranus models to the higher-frequency p modes.
Figures~\ref{fig.echelle}a-b show profiles of density, sound speed, and acoustic radius in a representative grid of interface models with $\ajump=2$ and $\rcore$ ranging from $0.1$ to $0.8\,\ru${, omitting for the moment any separate atmosphere polytrope}.
{(Here the acoustic radius
\begin{equation}
    \label{eq.acoustic_radius}
    t(r)=\int_0^r c_s^{-1}\,dr^\prime
\end{equation}
gives the sound crossing time from the planetary center to a radius $r$, and a complementary acoustic {\it depth} can be defined as $\tau(r)=\int_r^R c_s^{-1}\,dr^\prime$.)
}
Figure~\ref{fig.echelle}c shows the \'echelle diagram constructed from {these models'} $\ell=1$, $m=0$ mode spectra, in terms of the cyclic inertial frame frequency $\nu=\sigma/2\pi$.
The included modes span the interface mode through the p mode of order $n\approx26$.

This type of diagram (e.g., \citealt{1983SoPh...82...55G}) plots mode frequencies as a function of the frequency difference between modes of successive radial order, a convenient means of quickly identifying departures from the asymptotic constant spacing known as the ``large frequency separation'' $\Delta\nu$ given by Equation~\ref{eq.delta_nu}. 
Similar diagrams can be constructed from real (incomplete) mode spectra{, where the radial order is not known a priori,} by plotting {the observed frequencies modulo $\Delta\nu$. This large frequency spacing is also not known a priori but an informed guess can be made based on models, and its estimate can be refined based on the observed spacings.}

The frequency spacings {in Figure~\ref{fig.echelle}c} are imprinted with a clear periodic deviation controlled by the properties of the interface.
This can be understood as a result of a phase shift experienced by acoustic waves as they encounter the interface \citep{1994MNRAS.268..880R}, introducing a periodic modulation into the mode frequencies whose period is controlled solely by the acoustic radius of the jump (see also \citealt{1994A&A...283..247M}).
The amplitude of the modulation is controlled by the amplitude of the jump.
A sequence of measured p mode frequencies hence presents a unique avenue toward measuring the depth of any major density interface in Uranus's interior.
Taking the $\rcore=0.4\,\ru$ model (palest orange) as an example, Figure~\ref{fig.echelle}b yields a (diametric) core sound crossing time $2\tau_{\rm core}=2\int_0^{\rcore} c_s^{-1}\,dr=1500$ s, for a frequency $0.67$ mHz. 
Dividing by this model's large frequency separation $\Delta\nu=0.16$ mHz yields the equivalent radial order spacing $\Delta n\approx4$, in agreement with the {period obtained from the numerically} calculated spectrum in Figure~\ref{fig.echelle}c.

\subsection{Challenges of multiple density or sound speed interfaces}
\label{sec.doppler.complicated_echelle}

\begin{figure*}
    \begin{center}
        \includegraphics[width=\textwidth]{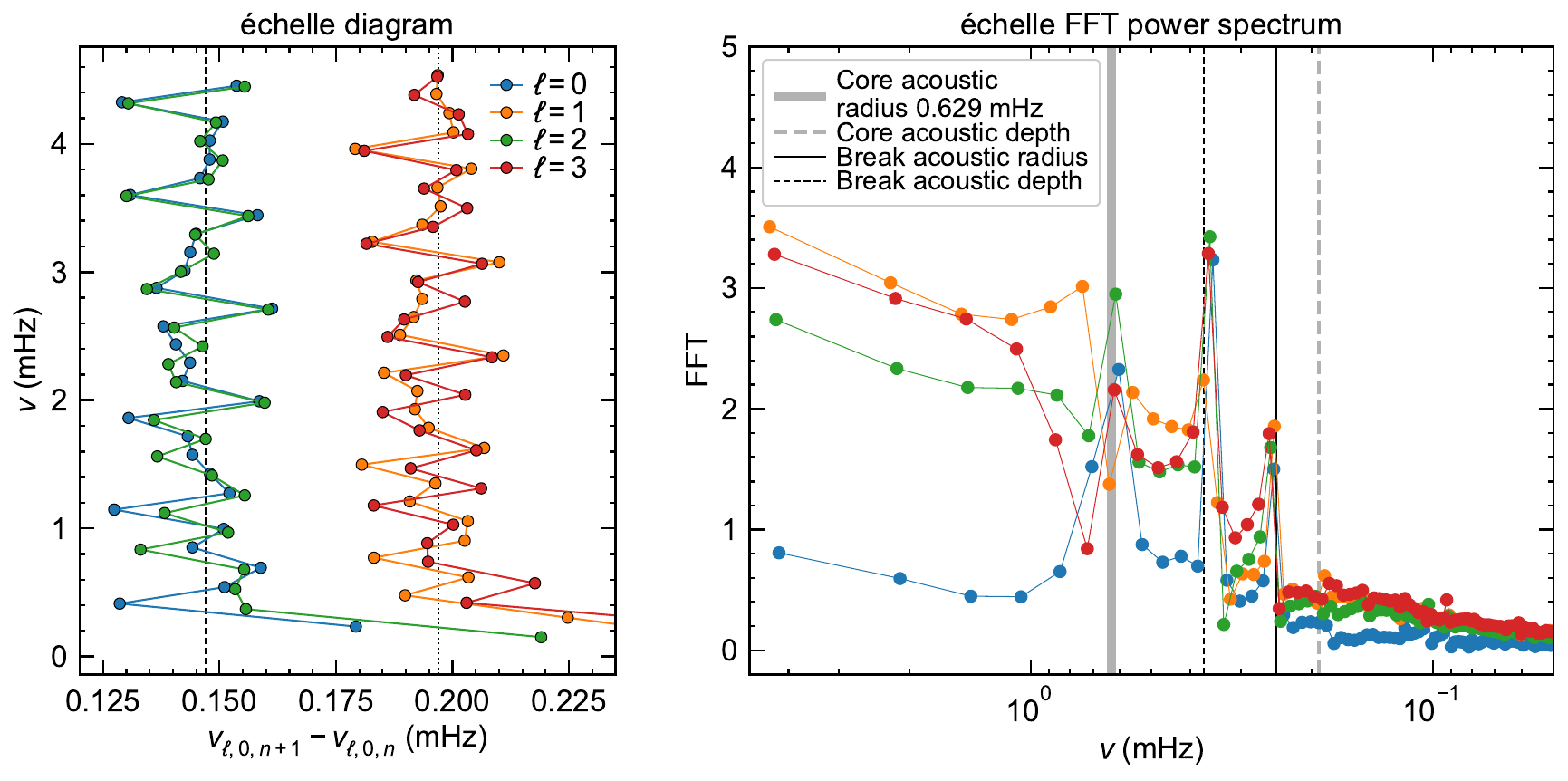}
        \caption{
            \label{fig.echelle_break_fft}
            {\textit{Left:} échelle diagram similar to Figure~\ref{fig.echelle}c, but for a model (the moderate central density interface model) including a discontinuous sound speed between the atmosphere and envelope polytropes. 
            $m=0$ modes for $\ell=0$--$3$ are shown, with the $\ell=1,3$ sequences offset horizontally by 0.05 mHz for clarity. 
            The vertical dashed line shows $\Delta\nu$; the dotted line is $\Delta\nu+0.05\ {\rm mHz}$.
            \textit{Right:} Naive FFTs of the frequency differences for each échelle sequence show a variety of peaks that may be difficult to disentangle if seen in the data. Vertical lines show the expected periodicities induced by the acoustic radius/depth of the core/break based on knowledge of this interior model.
            At face value the multiple structures at play would make this spectrum difficult to uniquely interpret; here techniques like the one discussed in Section~\ref{sec.doppler.phase_shifts} can be used.
            }
        }
    \end{center}
\end{figure*}

For clarity of demonstration, the models in Figure~\ref{fig.echelle} omit the atmosphere/envelope break usually included in our models. 
As a result, these models likely overestimate the density in Uranus's atmosphere {(see Section~\ref{sec.methods.models})}. 
Including a break {yields a more complicated, potentially more realistic} p mode spectrum, limiting the conclusions that can be drawn directly from a simple échelle diagram.
{
For example, Figure~\ref{fig.echelle_break_fft}a shows the $\ell=0-3$, $m=0$ échelle diagrams in our moderate $\rho_{\rm c}$ interface model with a core boundary $\rcore=0.43\,\ru$ and atmosphere/envelope break at $\rbreak=0.9\,\ru$ (see Figure~\ref{fig.models_of_interest}).
The break has introduced strong new periodicities into the frequency spacings, confounding their direct interpretation.
Figure~\ref{fig.echelle_break_fft}b shows a fast Fourier transform (FFT) of these frequency spacings, revealing 2-3 obvious periodicities\footnote{Note here the ``periods'' in question are in units of frequency (mHz) because the FFT is applied to a frequency spectrum and not a time series.}. 
The four vertical lines give the periods expected of the acoustic radius and depth for the core boundary and break, respectively, based on knowledge of the model's sound speed profile.
The two strongest local maxima evident for all $\ell$ correspond to the acoustic radius and depth expected for the break at $0.9\,\ru$.
The period corresponding to the acoustic radius of the core boundary is more subdued and strongest in the $\ell=0$ spectrum than the others.
The period corresponding to the acoustic \textit{depth} of the core boundary is not detectable.
Altogether, without the benefit of knowing Uranus's sound speed profile a priori, it would be difficult to arrive at a unique interpretation for an observed spectrum as complex as Figure~\ref{fig.echelle_break_fft}.
}
{In what follows we show how the frequencies of radial ($\ell=0$) and dipole ($\ell=1$) modes can be analyzed jointly to correctly locate the boundary in this more realistic case.}

\subsection{Insights from higher-order frequency differences}
\label{sec.doppler.phase_shifts}

\begin{figure}
    \begin{center}
        \includegraphics[width=0.5\textwidth]{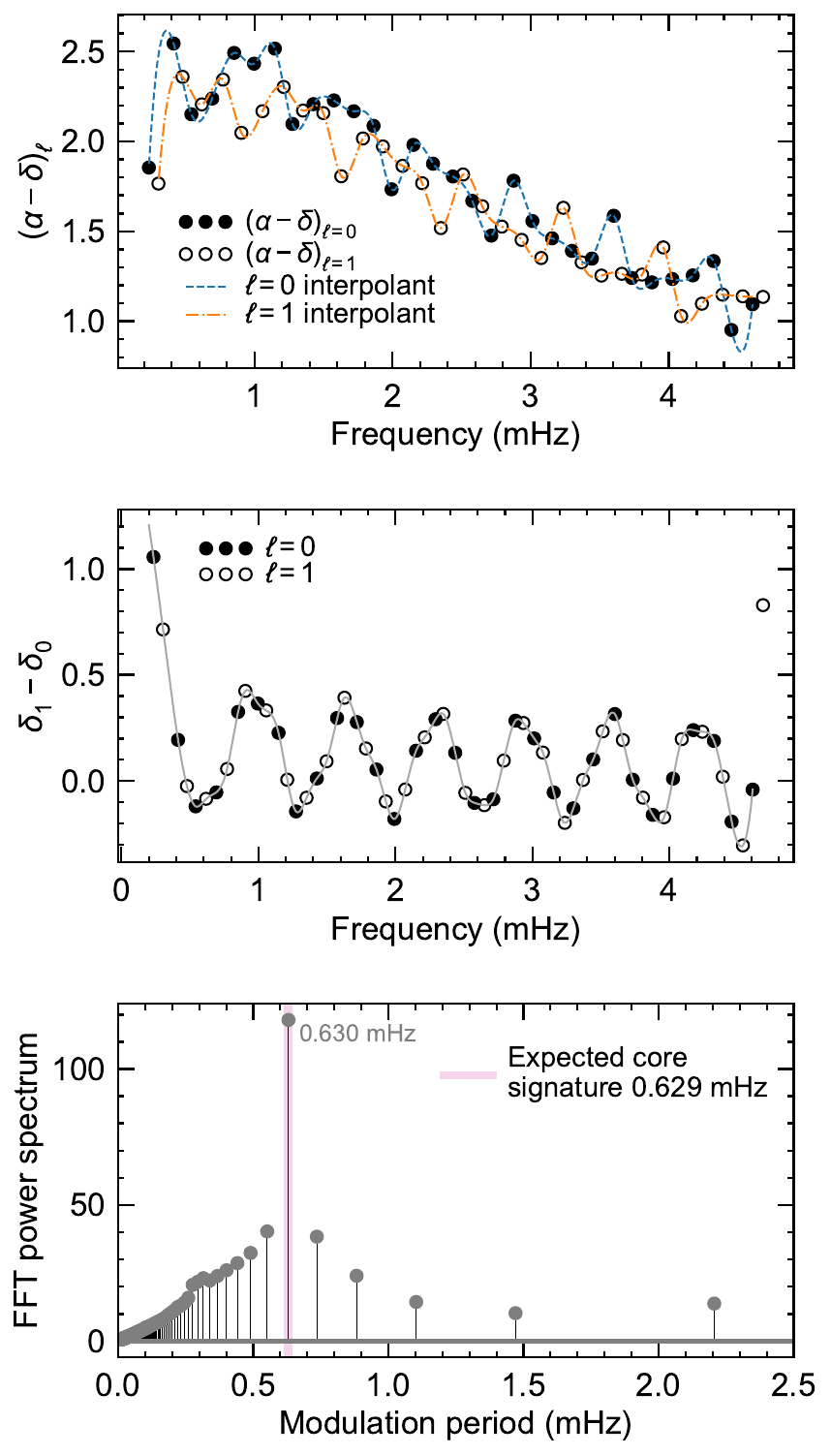}
        \caption{
            \label{fig.inner_phase_differences}
            An example of exploiting observations of modes of different $\ell$ to distinguish core structure from near-surface effects.
            \textit{Top:} The difference in outer and inner phase shifts as estimated from the $\ell=0,1$ spectra of the moderate $\rho_{\rm c}$ interface model; see Equation~\ref{eq.roxburgh_eigenfrequency} and discussion in text.
            Filled and open points show $\ell=0$ and $\ell=1$ respectively, and cubic spline interpolants are shown in the dashed curves.
            \textit{Middle:} The difference $(\alpha-\delta)_0-(\alpha-\delta)_1\simeq\delta_1-\delta_0$ calculated by differencing the interpolating functions. 
            \textit{Bottom:} An FFT of $\delta_1-\delta_0$ used to extract the period that remains in the inner phase shifts. The clear peak matches the {inverse} acoustic radius of the core calculated from the interior model.
        }
    \end{center}
\end{figure}

In the event that a measured p mode spectrum shows evidence for more involved near-surface effects, {the effects of deep and shallow features can be separated using higher-order frequency differences involving modes with different values of $\ell$, such as the small frequency separations (see, e.g., \citealt{1994MNRAS.267..297R}).
As one example of an application practical for Uranus, we condense some of the key results of \cite{2009A&A...493..185R}. Regarding the planet as split into inner and outer layers separated by a core boundary with acoustic radius $\tcore$ and acoustic depth $\taucore$ {(see Equation~\ref{eq.acoustic_radius})}, the solution to the oscillation equations obeys an ``eigenfrequency equation'' \citep{2000MNRAS.317..141R,2003A&A...411..215R}
\begin{equation}
    \label{eq.roxburgh_eigenfrequency}
    \pi\frac{\nu_{n,\ell}}{\Delta\nu} = \pi\left(n+\frac{\ell}{2}\right)+\alpha_\ell(\nu,\taucore)-\delta_\ell(\nu,\tcore)
\end{equation}
for integer $n$.
{While t}he inner phase shift $\delta_\ell$ and outer phase shift $\alpha_\ell$ could be calculated {numerically} from mode eigenfunctions given an interior model, {a model-independent result can be obtained by} recognizing that from a measured spectrum $\nu_{n,\ell}$ and an estimate for $\Delta\nu$, the difference in phase shifts $(\alpha-\delta)_\ell$ can be estimated directly. 
Figure~\ref{fig.inner_phase_differences}a shows these differences calculated from the $\ell=0,1$ frequencies for the same model as in Figure~\ref{fig.echelle_break_fft}.
The final ingredient is the theoretical result that the outer phase shift $\alpha_\ell$ is nearly independent of $\ell$.
This implies that differencing the two curves in Figure~\ref{fig.inner_phase_differences}a causes the outer phase shifts to cancel, yielding a signal $\delta_1-\delta_0$ (Figure~\ref{fig.inner_phase_differences}b) containing only a single strong periodicity representing the acoustic radius of the core boundary.
Figure~\ref{fig.inner_phase_differences}c shows that an FFT of $\delta_1(\nu)-\delta_0(\nu)$ recovers the correct location for the core boundary.
}
    
Finally we note that in models with a perfectly rigid core, the {lack of the central cavity eliminates} these phase shifts and hence {any} periodic signal in the p mode frequency spacings associated with the core. 
If Uranus's core is solid superionic H$_2$O as proposed by \cite{2021PSJ.....2..222S}, and assuming any solid body (e.g., torsional) oscillations supported by the frozen core do not couple with the p mode spectra, then barring any shallower discontinuities the spectrum resembles the asymptotic spectrum $\nu_{n+1}-\nu_n\approx\Delta\nu$ and the échelle diagram of Figure~\ref{fig.echelle}c becomes essentially featureless.
{However, in this case the large frequency spacing $\Delta\nu$ is likely to be drastically modified by the diminished acoustic cavity, and hence the value of the roughly uniform spacing in the observed frequencies itself becomes a diagnostic of the core size.}

Structure could be introduced into the échelle diagram by effects closer to the atmosphere (e.g., the atmosphere/envelope break included in earlier sections), but the aforementioned higher-order frequency differences and ratios thereof \citep{2003A&A...411..215R} can be exploited to isolate contributions from the core. 
Hence p mode seismology may be an unambiguous diagnostic of the state of the core, offering an opportunity to independently confirm---and more precisely quantify---any conclusions drawn from new measurements of Uranus's tidal dissipation and its tidal Love number $k_{22}$ \citep{2021PSJ.....2..222S,2024PSJ.....5..116P}.

\section{Discussion}\label{sec.discussion}
A large set of frequencies that can be used to decode Uranus's internal structure are potentially lying in wait. 
However, 
{
    there is no guarantee that Uranus's normal modes are excited to observable amplitudes.
    Even in Saturn for which the data are extensive, the processes responsible for exciting and dissipating the observed modes remains an open question \citep{2018Icar..306..200M,2019ApJ...881..142W} that we do not attempt to resolve here.
    Nonetheless, Saturn's f modes are excited and apparently coherent over a timescale of at least decades \citep{2022PSJ.....3...61H}.
    As already mentioned, Saturn and Jupiter also both show indirect evidence for higher frequency (p mode) seismicity as glimpsed in radio tracking of the Cassini and Juno trajectories \citep{2020PSJ.....1...27M,2022NatCo..13.4632D}.
    {We show in Appendix~\ref{app.ring_mode_energy} that a Uranian f mode can generate a similar magnitude of Lindblad forcing potential to a known resonance between the satellite Ophelia and the $\gamma$ ring, provided that the f mode has an energy on the order of $10^{24}\ {\rm erg}=5\times10^{-18}~GM_{\rm U}^2/\ru$. For comparison, Saturn ring seismology indicates typical f-mode energies $\sim2\times10^{25}~{\rm erg}~5\times10^{-18}\ GM_{\rm S}^2/R_{\rm S}$ \citep{2014Icar..242..283F}. Despite the similar fractional energies in the two cases, there is no reason a priori to expect the same mode excitation and dissipation mechanisms to operate in both planets.}
    It is {also} possible that these modes are excited to greater amplitudes at Uranus than at Saturn or Jupiter.
    {Uranus's weak intrinsic flux might invite skepticism in this regard, but the possibility of a hot, convective interior insulated from the cold atmosphere by a thermal boundary (e.g., \citep{2016Icar..275..107N,2012A&A...540A..20L}) suggests that a large reservoir of fluid kinetic energy cannot be ruled out.}
    Altogether, there is reason to be optimistic that Uranus's oscillations will be detectable by one means or another, {especially given} the high-reward nature of the science that these techniques could enable.
    {More Earth-based seismology measurements of Jupiter and Saturn, and renewed efforts to understand the amplitudes of Saturn modes observed in Saturn's rings, would help to inform expectations for mode amplitudes at Uranus or Neptune.}
}

Here we have considered the rings and Doppler imaging of Uranus's surface as two windows through which we might gather complementary data about Uranus's oscillation spectrum and hence its confounding internal structure. Ring and Doppler imaging seismology each come with their intrinsic opportunities and challenges.

Saturn seismology by Cassini was successful despite not being a part of the mission design. Ring seismology has the tremendous advantage of being possible without dedicated instrumentation, achievable in its most basic form from imaging data {revealing $m$-fold periodic brightness perturbations or embedded density or bending waves \citep{2009Icar..202..260H,2011epsc.conf.1224S,2023DDA....5440202H,2023LPICo2808.8017H}}. {Stellar occultations would be preferable, but not strictly necessary.}
Hence {it is likely that} these low-cost observations will proceed to some degree so long as an orbiter flies, without the need for additional motivation.


Our calculations here are intended to lay out some guidance as to how features in Uranus's largely unconstrained interior might manifest in the rings, building on \cite{2022PSJ.....3..194A} by studying a more exhaustive statistical sample of interior structures (Section~\ref{sec.methods}) {and performing retrievals (Section~\ref{sec.rings})}. In particular we consider the possibilities of a fluid or solid core as well as composition gradients, some combination of which are likely to be present in the interior given Uranus's vanishingly small heat flux \citep{2016Icar..275..107N,2021PSJ.....2..222S,2020A&A...633A..50V}. 
{On this note, Uranus's interior is likely to include regions of superadiabatic temperature stratification \citep[e.g.,][]{2024A&A...684A.191N}, an effect we have not explicitly considered here. 
This superadiabaticity may manifest as either double-diffusive convection ({possibly} in the form of convective layering{, e.g., \citealt{2007JFM...577..251R,2012ApJ...750...61M,2024ApJ...975L...1F}}) or {diffusive/radiative} transport.
{In the opaque convective or semiconvective regions of {gas giants like Jupiter and Saturn}, the thermal superadiabaticity $\nabla-\nabla_a$ is a minor correction to the buoyancy compared to the dominant composition term $B$ (see \citealt{2021NatAs...5.1103M}). 
{It is not clear yet to what degree this property holds in the uncertain, heavy element dominated conditions of the Uranian interior. Additionally}, radiative regions spawned by deep atmospheric opacity windows \citep{2023A&A...680L...2H,2024ApJ...967....7M}, inhibited convection in the cloud layers \citep{2017A&A...598A..98L,2017Icar..297..160F,2021PSJ.....2..146M}, H/He immiscibility layers \citep{2024arXiv240913895M}, or H$_2$O/H$_2$ immiscibility layers (\citealt{2021PSJ.....2...64B,CanoAmoros2024,2024arXiv240704685G}; but see \citealt{2015ApJ...806..228S}) could present significant features in the thermal buoyancy that should be addressed by future work in giant planet seismology.}
}

We have also introduced a framework for directly fitting interior models to observed resonances (Section~\ref{sec.rings}). We have demonstrated that extracting a single mode's frequency and azimuthal pattern number $m$  using ring seismology {would allow us to eliminate a large fraction of the models permitted by zonal gravity.} 
{Some ambiguity survives, }especially as a function of which radial order mode is held responsible {for a given ring pattern}. 
In ambiguous cases the detection of a second mode would {largely break} the degeneracy, two points serving to `calibrate' the inherently highly structured ring spectrum (see, e.g., \citealt{2023PSJ.....4...59M} Figure 12). {Nonetheless we caution that, in light of the rich complexity of Uranus interior physics, our simplified parameterization in terms of modified polytropes (Section~\ref{sec.methods.models}) likely cannot cover the full space of possible interior structures. Importantly, it does not capture details like the temperature dependence of the density. It is therefore possible that the analysis of normal mode detections from a future UOP will encounter degeneracies that we have not fully contended with here. On a more optimistic note, these degeneracies may be mitigated by complementary information from new measurements of Uranus's gravitational and magnetic fields.}

{
    If a low-degree f mode, ideally $_2^2$f, resonates with any of the narrow rings, its detection would place an important new constraint on Uranus's core extent and density (see Section~\ref{sec.rings.fits}). However, this outcome relies on a relatively extended rigid core ($\rcore/\ru\gtrsim0.4$ if the $_2^2$f OLR is near the 6 ring, as an example; see Figure~\ref{fig.6ring_99f_fit_scatter}). For very compact cores, as in the maximum $\rho_{\rm c}$ rigid core model where a $0.6$ Earth mass core resides within $\rcore/\ru\approx0.17$ (Figure~\ref{fig.models_of_interest}), even the most deeply reaching $_2^2$f mode has close to zero sensitivity to the core. Hence, compact, sharply defined cores may evade detection in ring seismology. The best tool for revealing the boundary of such a core is Doppler imaging, by virtue of the p mode frequency spacings' sensitivity to core boundaries near $\rcore/\ru\approx0.2$ or deeper (see Section~\ref{sec.doppler.echelle} and Figure~\ref{fig.echelle}).
}

Ring seismology may appear to have the drawback of relying on chance overlaps between planet mode frequencies and sharp ring features, but the connection between the two might ultimately be causal in at least some cases, especially considering the {abundance of narrow features versus the} dearth of known ring-satellite resonances in the rings of Uranus. The odds of finding features associated with Uranus's interior are further helped by the fact that that there are numerous additional narrow features visible in high-phase imaging of the rings \citep{2021PSJ.....2..107H}, supplementing the named rings that we have considered in detail in Section~\ref{sec.rings} (see \citealt{2022PSJ.....3..194A}, particularly their Section~5.2). Patterns forced by Uranian normal modes may be observable in spiral density waves embedded in narrow, dense rings, or in perturbations to the shape of ring edges; see the discussion of detection methodologies in \cite{2022PSJ.....3..194A}. They discuss the fact that even Fresnel-limited occultations from Earth probably would not {attain} sufficient resolution to characterize new waves, although at the lowest azimuthal wavenumbers like $m\sim2$ this may be more achievable.

The most transformative application of Uranian seismology would involve multimodal observations, combining ring seismology with another technique with sensitivity to different parts of the planet's spectrum. Two avenues we see are Doppler imaging and direct gravitational seismology by an orbiter. 

Doppler imaging would be most sensitive to p modes because their higher frequencies imply larger radial velocity amplitudes than the f, g, or interface modes at comparable mode energies. These frequencies are inaccessible to ring seismology, as are zonal $m=0$ and retrograde $m<0$ modes. Hence Doppler imaging has the potential to expose an extensive list of mode frequencies that can be readily interpreted using mature techniques from helioseismology and asteroseismology (see \citealt{2002RvMP...74.1073C}). In particular p modes are sensitive to features in the first adiabatic index (i.e., the adiabatic sound speed), and their higher order radial eigenfunctions have increased sensitivity to short wavelength features of the planetary interior. As an example, in Section~\ref{sec.doppler} we have shown that sequences (not necessarily unbroken ones) of $\sim$ten $\ell=1$ p mode frequencies may be sufficient to locate a core boundary in Uranus if the boundary is sharp. The presence or absence of a periodic trapping signature in the p mode frequencies would be diagnostic of the core's material phase, i.e., frozen or fluid, independent of constraints that may be gleaned from Uranus's tidal response \citep{2021PSJ.....2..222S,2023PSJ.....4..241N,2024PSJ.....5..116P}.

Doppler imaging from an orbiter has practical limitations that warrant in-depth study.
These observations typically target one or more solar absorption lines, making use of the tendency {of the planet's oscillating surface} to shift reflected light toward or away from the line center.
{For MOF instruments with narrow, fixed bandpasses mandated by their filter design (e.g., PMODE; \citealt{2022FrASS...968452S}), this puts an upper limit on the spacecraft-planet radial velocity that can be tolerated during Doppler imaging measurements.
Hence for an eccentric orbit, observations from such an instrument might need to be concentrated around apoapse when the absorption line is close to its rest frame wavelength.
Interferometers (e.g., JIVE; \citealt{2024PSJ.....5..100S}) can sidestep this limitation entirely. 
Observations at apoapse may be preferred in any case due to reduced competition with other instruments and the opportunity for continuous observations to build a long baseline time series.}
More detailed mission studies will need to consider the planetary phase angle achievable near apoapse for realistic orbits.

Gravitational seismology (\citealt{2020RSPTA.37890475F}; Friedson et al.~2025) involves the search for oscillation modes through their influence on the trajectory of an orbiter, specifically targeting the `anomalous' accelerations that confounded measurements of Saturn's static gravity field \citep{2019Sci...364.2965I}. Accelerations of this type in Cassini data were attributed to Saturn normal modes \citep{2020PSJ.....1...27M}, and similarly to Jupiter normal modes in Juno data \citep{2022NatCo..13.4632D}. The orbit design of a future Uranus mission may benefit from considering the prospect of gravitational seismology, and methods should be devised for translating these data into concrete measurements of frequency, $\ell$, and $m$.

\section{Conclusion}\label{sec.conclusion}
The field of normal mode seismology for spheroidal fluid bodies is mature thanks to the abundance of data for the Sun and other stars (see \citealt{2002RvMP...74.1073C,2010aste.book.....A}).
These approaches have long been pursued in the domain of giant planets (e.g., \citealt{1976Icar...27..109V,1993A&A...267..604M,1999P&SS...47.1211G}) and are coming to fruition thanks to ring seismology of Saturn by Cassini \citep{1993Icar..106..508M,2013AJ....146...12H} and continued refinement of ground-based observations of Jupiter (\citealt{2022FrASS...968452S,2024PSJ.....5..100S}).
Here we have merely scratched the surface of these techniques as they might be applied to Uranus from a future orbiter.

{Doppler imaging from a Uranus orbiter would likely generate the most information suitable for Uranian seismology, provided that internal processes excite Uranus's p modes to observable amplitudes. The amplitudes that would be necessary depend on an intersection of instrumental and mission design considerations that have yet to be studied in detail. This line of study is warranted given the high sensitivity of the p modes to compositional interfaces in Uranus's interior, the characterization of which could unravel the puzzles of Uranus's internal heat flow, magnetic field generation, and thermodynamics.}

{Ring seismology is fundamentally sensitive to lower frequency modes, including f, g, and interface modes. The detection of such a mode with this method relies on a mode's frequency coinciding with natural frequencies of ring orbits, an alignment that is far from guaranteed in the sparse rings of Uranus. However, the information that can be gleaned about Uranus's deep interior structure and/or rotation period from even one or two such detections motivates the relatively low-cost observations (stellar occultations or ring imaging) needed to search for these precious traces of Uranus's dynamic gravity field.}

\section*{Acknowledgments}
    The authors thank the two anonymous reviewers for their thoughtful suggestions.
    This work benefited from ideas and advice from the participants of the September 2023 workshop ``Determining the Interior Structure of Uranus" organized by the W.M. Keck Institute for Space Studies.
    C.M.'s research was supported by an appointment to the NASA Postdoctoral Program at the Jet Propulsion Laboratory, administered by Oak Ridge Associated Universities under contract with NASA.
    C.M. thanks N. Nettelmann, B. Militzer, R. French, P. Nicholson, S. Müller, B. Idini, and R. Helled for helpful conversations.
    Pre-Decisional Information for planning and discussion purposes only. 
    The research described in this paper was carried out at the Jet Propulsion Laboratory, California Institute of Technology, under a contract with the National Aeronautics and Space Administration (80NM0018D0004). 
    \copyright 2025 California Institute of Technology, all rights reserved. Government sponsorship acknowledged.

\bibliography{references}
\bibliographystyle{psj}

\appendix
\allowdisplaybreaks
\section{Mode energy to yield an observable ring response}
\label{app.ring_mode_energy}
{
    {To assess whether it is plausible for Uranus f modes to drive observable signatures in the rings, we look to} a known satellite resonance {that drives an observable ring response}.
\cite{2024Icar..41115957F} characterized an $m=6$ mode on Uranus's $\gamma$ ring driven by the Ophelia 6:5 inner Lindblad resonance. 
The ring mode is detected in occultations at both the inner and outer edges of the $\gamma$ ring, its measured amplitude yielding an estimate for Ophelia's mass.
For comparison to forcing by a planet mode, it is useful to note the resonance's effective perturbing potential $\Psi\approx 7\times10^3\ {\rm cm}^2\ {\rm s}^{-2}$ (see \citealt{1984prin.conf..513S} for a review of linear density wave theory). 
Meanwhile, the effective perturbing potential of an OLR with a planet mode is a function of the mode's gravitational potential perturbation $\Phi^\prime$ and its radial gradient, evaluated at the Lindblad resonance (see, e.g., Appendix B of \citealt{2021PSJ.....2..198D}).
Demanding that a $\ell=m=2$ mode have the same magnitude of forcing potential as the Ophelia resonance,
\begin{equation}
    \label{eq.mode_ophelia_potential_equivalence}
    \Psi[_2^2{\rm f\ OLR}]\sim\Psi[{\rm Ophelia\ 6:5\ ILR}]=7\times10^3\ {\rm cm}^2\ {\rm s}^{-2},
\end{equation} 
implies a nondimensional mode amplitude of $A\sim3\times10^{-9}$, mode energy $\sim10^{24}$ erg, surface radial displacement amplitude $\sim30\ {\rm cm}$, and surface radial velocity amplitude $\sim4\times10^{-3}\ {\rm cm\ s}^{-1}$. Low order g mode energies in our models are typically comparable with the f modes within an order of magnitude.
}

{
{This} mode energy is $\sim5\%$ that of a typical Saturn f mode given information from the rings \citep{2014Icar..242..283F,2019ApJ...881..142W}. {For comparison, }Uranus's binding energy is approximately $5\%$ of Saturn's. We conclude that, whatever mechanisms might drive Uranus's normal modes, {it is energetically reasonable for the f modes to generate ring features that rival the ones generated by natural satellites. Hence, t}he types of observations proposed by~\cite{2022PSJ.....3..194A} and in Section~\ref{sec.rings} are {worthy of pursuit}.
}

\section{Wind model and thermal wind balance}
\label{app.tgwe}

\begin{figure*}
    \begin{center}
        \includegraphics[width=0.8\textwidth]{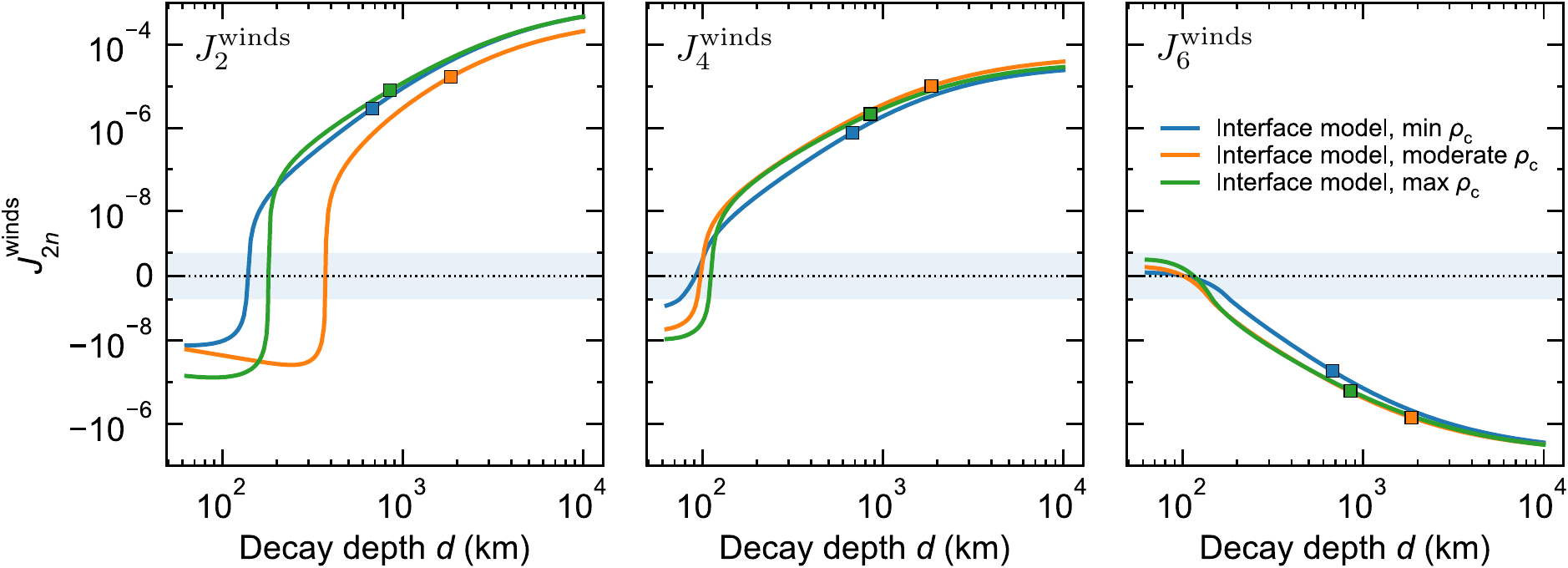}
        \caption{
            Perturbations to the zonal gravity moments $J_{2n}$ induced by the wind according to thermo-gravitational wind balance, assuming the observed wind speeds are constant on cylinders before applying exponential decay as a function of radial depth from the surface. 
            Shown as a function of the $e$-folding depth $H$.
            {Square symbols designate the true retrieved depth and associated harmonics $J_{2n}^{\rm winds}$ for each model.
            The vertical axis switches to linear scale between $-10^{-9}$ and $+10^{-9}$ (shaded region).
            }
        \label{fig.tgwe_djn}
        }
    \end{center}
\end{figure*}

\begin{figure*}
    \begin{center}
        \includegraphics[width=\textwidth]{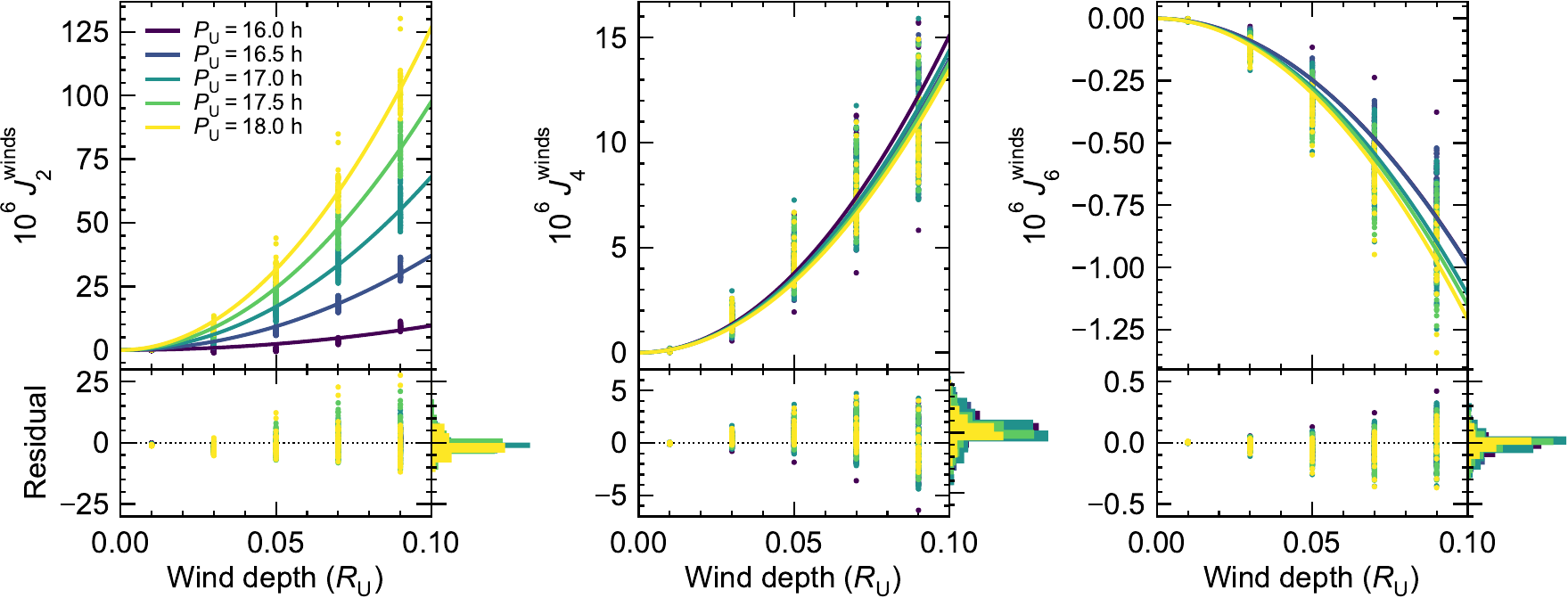}
        \caption{
            \label{fig.dj2n_fits}
            Polynomial least-squares fits to the wind-induced components of $J_2$, $J_4$, and $J_6$ as a function of wind depth and bulk spin period. These reflect the results of full TGWE calculations (see Section~\ref{sec.methods.models.constraints}) for $\sim1000$ interface models fit to $J_2$ and $\rho_1$ alone. Bottom panels with histograms to the right show that the magnitude of the residuals are typically $\lesssim5$ ppm in $J_2$, $\lesssim1$ ppm in $J_4$, and $\lesssim0.1$ ppm in $J_6$.
        }
    \end{center}
\end{figure*}

Figure~\ref{fig.tgwe_djn} shows $J_{2n}^{\rm winds}$ obtained from solutions to thermal wind balance, as a function of the assumed decay depth $d$ of the winds. These models are the end member interface models shown in Figure~\ref{fig.models_of_interest}.
Our polynomial fit to $J_{2n}^{\rm winds}(d,\Omega)$ is shown in Figure~\ref{fig.dj2n_fits}. The top panels show full TGWE values (points) and polynomial fits (curves). The bottom panels show the fit residuals, with their histograms along the right axes. 
Of $\mathcal O(10^4)$ randomly chosen models from a sample of rigidly rotating interface models fit to $J_2$ and $\rhoone$, only models with sampled $\pu$ falling within narrow bins (width $\approx 2$ minutes) around predetermined values $\pu=16.0$, 16.5, 17.0, 17.5 and 18.0 h are retained, leaving 1,810 models total. 
For each model, flow profiles across a grid of wind depths are considered, and the resulting $J_{2n}^{\rm winds}$ {are} tabulated as a function of $\pu$ and $d$. We construct quadratic fits to $J_{2n}^{\rm winds}(d|\pu)$. These fits and their residuals are shown in Figure~\ref{fig.dj2n_fits}. The residual random scatter reflects the weaker dependence the wind-induced moments have on other model parameters not accounted for in the fit, leaving RMS errors $<5$ ppm for $J_2^{\rm winds}$, $<2$ ppm for $J_4^{\rm winds}$, and $<0.3$ ppm for $J_6^{\rm winds}$. Models evaluated during MCMC use these fitting functions and linear interpolation in $\pu$ to evaluate $J_{2n}^{\rm winds}$ for arbitrary $\pu$ and $d$.

\section{Parameter Estimation}
\label{app.parameter_estimation}
{Section~\ref{sec.methods.parameter_estimation} briefly described our procedure for fitting models to data. Here we give more technical details and show the posterior probability distributions for the baseline samples fit to $J_2$, $J_4$, and $\rhoone$.}
Since estimates for $J_2$ and $J_4$ rely principally on the dynamics of Uranus's relatively closely packed rings, these quantities are highly correlated. We incorporate Jacobson's $J_2-J_4$ correlation of $0.981$ (see \citealt{2024Icar..41115957F}) into our multidimensional likelihood function, leading to the strong covariance that can be observed in Figure~\ref{fig.j2n_scatter_gradient_break}. Any covariance between the $J_{2n}$ and $\rhoone$ is neglected. 

All models use $J_{2n}$ centroids from \cite{2014AJ....148...76J}. Baseline models also adopt Jacobson's uncertainties and strong $J_2$--$J_4$ covariance, with the exception of Section~\ref{sec.rings.fits_improved_gravity} which considers improved information on zonal gravity from an orbiter \citep{2024PSJ.....5..116P}. Model parameters are summarized in Table~\ref{tab.parameters} and the prior probability is assumed uniform in the allowed parameter space. Sampling is achieved using {\texttt emcee}\footnote{\url{https://emcee.readthedocs.io/en/v3.1.4}} \citep{2013PSP..125..306F}. 

\begin{figure*}
    \begin{centering}
        \includegraphics[width=\columnwidth]{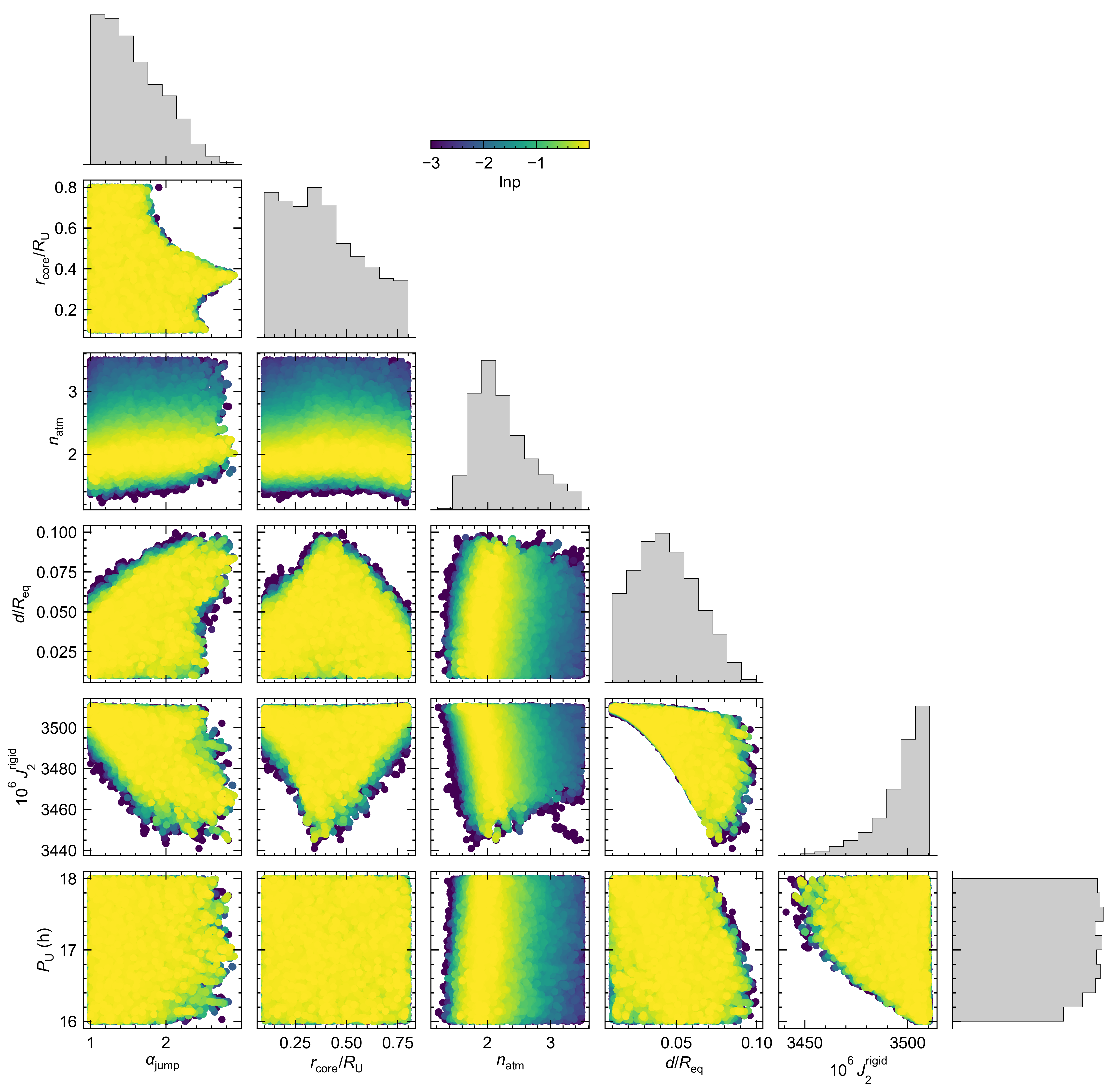}
        \caption{
            \label{fig.corner_jump_break} 
            Corner diagram for interface models constrained by gravity and $\rhoone$.
            }
    \end{centering}
\end{figure*}
\begin{figure*}
    \begin{centering}
        \includegraphics[width=\columnwidth]{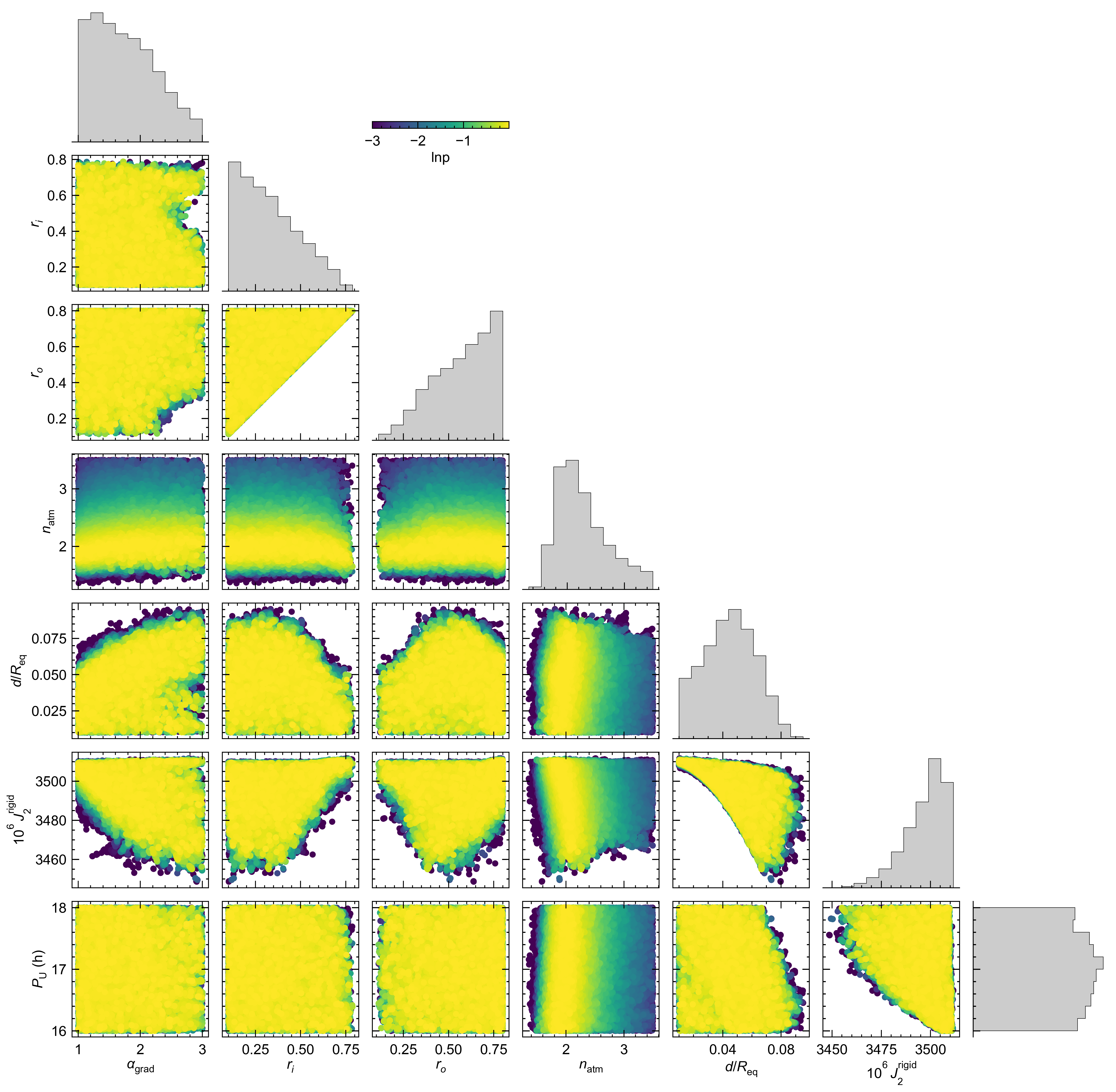}
        \caption{
            \label{fig.corner_gradient_break} 
            Corner diagram for gradient models constrained by gravity and $\rhoone$.
            }
    \end{centering}
\end{figure*}

Figure~\ref{fig.corner_jump_break} shows the posterior probability distributions obtained for the baseline sample of interface models. 
Figure~\ref{fig.corner_gradient_break} shows that for the gradient models.
{In each case only 1 model is plotted for every 10 models in the sample.}
{These diagrams illustrate the vastly degenerate solution space allowed by the present constraints on Uranus's gravity field.}

\section{Sampling Strategy for Precise Gravity}
\label{app.rejection_sampling}

Section~\ref{sec.rings.fits_improved_gravity} focuses on the question of whether ring seismology constraints are of value even in light of new information expected from the UOP determination of Uranus's gravity field.
Even a conservative estimate for the precision on gravity moments expected from UOP (\citealt{2024PSJ.....5..116P}; see Section~\ref{sec.methods.models.constraints}) leads to difficulty when using our standard fitting process. 
The sharply peaked likelihood function yields impractically low acceptance fractions, even when we consider alternative proposal distributions offered by {\texttt emcee}.
In this case we opt to use rejection sampling, a simpler, classical sampling algorithm (see \citealt{8d981afb-374e-301d-8157-b0d677d4bcbf}) wherein the probability of accepting a given Monte Carlo step is independent of the previous step. 

The simplest version of this algorithm would draw a parameter vector $\theta_i$ uniformly distributed over the prior volume, run a model to evaluate $\ln L(\theta_i)$, and draw a random number $x_i$ uniformly distributed within $(0,1)$. Then the model is accepted into the sample if $\ln L(\theta_i) > Cx_i$ for a predetermined constant $C>0$; otherwise it is rejected.
Our large parameter space implies that drawing uniformly from the allowed prior volume (see Table~\ref{tab.parameters}) would yield vanishingly few acceptable models. 
Instead we begin with a relatively permissive sample of models loosely constrained by $J_2$ and $J_4$, generated by our normal sampling process but using a modified likelihood $\tilde L$ with Gaussian errors equal to the Parisi error bars inflated by a factor of 5.
For each model in this initial sample, the final likelihood $L$ is calculated using the true Parisi gravity errors, and the model is accepted or rejected based on whether the ratio $L(\theta_i)/\tilde L(\theta_i)$ exceeds the uniformly distributed $x_i\in(0, 1)$.
The process is then repeated with an alternate final likelihood that folds in an additional seismic constraint in the form of a $_2^2f$ OLR on the 6 ring, calculated assuming a rigid core as in Section~\ref{sec.rings.fits}. Starting from $3\times10^5$ models in the permissive sample, the stronger gravity constraints yielded 80,570 accepted models. 
The stronger gravity constraints paired with the ring seismology constraint yielded the 1,207 models that are shown in Figure~\ref{fig.rejection_resampling_fit_mode}.

\end{document}